\documentclass[11pt]{article}
\usepackage{latexsym,amsmath,amssymb,theorem,epsfig,hyperref}

\usepackage{cite}

\numberwithin{equation}{section}

\makeatletter
\DeclareFontFamily{OMX}{MnSymbolE}{}
\DeclareSymbolFont{MnLargeSymbols}{OMX}{MnSymbolE}{m}{n}
\SetSymbolFont{MnLargeSymbols}{bold}{OMX}{MnSymbolE}{b}{n}
\DeclareFontShape{OMX}{MnSymbolE}{m}{n}{
<-6>  MnSymbolE5
<6-7>  MnSymbolE6
<7-8>  MnSymbolE7
<8-9>  MnSymbolE8
<9-10> MnSymbolE9
<10-12> MnSymbolE10
<12->   MnSymbolE12
}{}
\DeclareFontShape{OMX}{MnSymbolE}{b}{n}{
<-6>  MnSymbolE-Bold5
<6-7>  MnSymbolE-Bold6
<7-8>  MnSymbolE-Bold7
<8-9>  MnSymbolE-Bold8
<9-10> MnSymbolE-Bold9
<10-12> MnSymbolE-Bold10
<12->   MnSymbolE-Bold12
}{}

\let\llangle\@undefined
\let\rrangle\@undefined
\DeclareMathDelimiter{\llangle}{\mathopen}%
               {MnLargeSymbols}{'164}{MnLargeSymbols}{'164}
\DeclareMathDelimiter{\rrangle}{\mathclose}%
               {MnLargeSymbols}{'171}{MnLargeSymbols}{'171}
\makeatother

\def\a{\alpha}
\def\b{\beta}

\def\e{\epsilon}

\def\k{\kappa}
\def\l{\lambda}
\def\m{\mu}
\def\n{\nu}

\def\G{\Gamma}

\def\L{\Lambda}

\def \N  {{\cal  N  }}

\def\gsim{ \lower .75ex \hbox{$\sim$} \llap{\raise .27ex \hbox{$>$}} }
\def\lsim{ \lower .75ex \hbox{$\sim$} \llap{\raise .27ex \hbox{$<$}} }
\def\be{\begin{equation}}
\def\ee{\end{equation}}
\def\bea{\begin{eqnarray}}
\def\eea{\end{eqnarray}}

\def \ha {{1 \ov 2}}

\def \del{\partial}
\def \a {\alpha}
\def \aa {{\a'}}
\def\ov{\over}
\def \ci {\cite}
\def \n {{\rm n}}

\def \foot {\footnote}

\def\la{\label}\def\foot{\footnote}
\newcommand{\rf}[1]{\eqref{#1}}

\def \no {\nonumber}

\def \adss {AdS$_5 \times S^5\ $}

\def \a {\alpha }

\def \l {\lambda}
\def\foot{\footnote}

\def \adss {AdS$_5 \times S^5~$ }
\def \fo {\tfrac{1}{4}}
\def \ov {\over}
\def \ci  {\cite}
\def \te {\textstyle}

\def \ed {\end{document}}
\def \F {{\rm F}}

\def \xx {{\rm x}}

\def \vx {\vec{x}}

\def \fo {{1\ov 4}}\def \n {\nu}
\def \la {\label} \def \fo {{1\ov 4}}
\def \ed   {\end{document}}
\def \beg  {\begin{equation}}
\def \eeg {\end{equation}}
\def \tet {\textstyle}
 
\def \DD {{\rm D}}
\def \TT  {{\rm T}_2 } 
\def \ep {\epsilon}

\def \RR {{\rm R}} \def \iffa {\iffalse} \def \aa {{\rm a }}  
\DeclareMathOperator{\Tr}{Tr}
\DeclareMathOperator{\tr}{tr}
\DeclareMathOperator{\vol}{vol}
\def \rr {{\rm r}}
\def \OO {{\cal O}}

\def \kk {\kappa}\def \ka {\kappa} \def \L  {\Lambda}
\def \rT {{\rm T}}

\def \tt {{\rm t}}
\def \half {\tfrac{1}{2}}

\def  \adsss  {AdS$_7\times S^4\ $}
\def  \adssp  {AdS$_5\times S^5\ $}
\def \adss {AdS$_7\ $}
\def \adst   {AdS$_3\ $}

\def \N   {{\cal N}}

\def \half {\tfrac{1}{2}}
\def \rC {{\rm C}}
\def \ze {\chi}

\usepackage{color}

\begin{document}

\vspace{-7cm}



\vspace{-5cm}

\title{
\begin{tabbing}
\hspace*{9cm} \=  \kill 
\> {\small \ \ \ \ \ \ \ \ \ \ \ \ \ \ \ \ \ \ \ \ \ \ \ \ \ \ \ \ \ \ \ \ \ \   Imperial-TP-AT-2020-02  } 
\end{tabbing}
\vspace{2cm}
{ Defect CFT in the 6d (2,0) theory\\
 from M2 brane dynamics in  AdS$_7 \times S^4$
 }
} 
\date{}
\author{Nadav Drukker$^a$,\, Simone Giombi$^b$,\, Arkady A. Tseytlin$^{c,}$\footnote{Also at Lebedev Institute and ITMP, Moscow State University.},\, Xinan Zhou$^d$
\\[0.5em]
{\it \small $^a$Department of Mathematics, King's College London, The Strand, London WC2R 2LS, U.K.}
\\[0.3em]
{\it \small $^b$Department of Physics, Princeton University, Princeton, NJ 08544, U.S.A.}
\\[0.3em]
{\it \small $^c$The Blackett Laboratory, Imperial College, London SW7 2AZ, U.K.}
\\[0.3em]
{\it \small $^d$Princeton Center for Theoretical Science, Princeton University, Princeton, NJ 08544, U.S.A.}
}

\maketitle
\begin{abstract}
Surface operators in the 6d (2,0) theory at large $N$ have a holographic 
description in terms of M2 branes probing the AdS$_7 \times S^4$ M-theory background. The most symmetric, $1\ov  2$-BPS, operator is defined over a planar or spherical surface, and it preserves a 2d superconformal group. This includes, in particular, an $SO(2,2)$ subgroup of 2d conformal transformations, so that the surface operator may be viewed as a conformal defect in the 6d theory. The dual M2 brane has an AdS$_3$ induced geometry, reflecting the 2d conformal symmetry. 
Here we use the holographic description to 
extract the defect CFT data associated to the surface operator.  
The spectrum of transverse fluctuations of the M2 brane is found to be  in one-to-one correspondence with a protected multiplet of operator insertions on the surface, which includes the displacement operator. 
We compute the one-loop determinants of fluctuations of the M2 brane, and extract the conformal anomaly coefficient of the spherical surface to order $N^0$.
We also briefly discuss the RG flow from the non-supersymmetric to the $1\ov  2$-BPS  defect  operator, and its consistency with a  ``$b$-theorem" for the defect CFT. 
Starting with the M2 brane action, we then use AdS$_3$ Witten diagrams to compute the 4-point functions of the elementary bosonic insertions on the surface operator, and extract some of the defect CFT data from the OPE. 
The 4-point function is shown to satisfy superconformal Ward identities, and we discuss a related subsector of ``twisted" scalar insertions, whose correlation functions are constrained by the residual superconformal symmetry.  
\end{abstract}

\maketitle
\flushbottom

\newpage

\tableofcontents

\setcounter{footnote}{0}

\def \xxx {z}
\def \xxxx {{\rm z}}
\def \ze {\chi}

\def \zz {{v}}

\section{Introduction  and summary}


Non-local operators are an important class of observables in conformal field theories in various dimensions. When they are defined over planar or spherical submanifolds, they may preserve a subgroup of the conformal symmetry of the ``bulk" CFT, and are often referred to as conformal defects. 
Using the AdS/CFT duality, one may develop a strong-coupling perturbation theory approach to the computation of their expectation values
and correlation functions of local operators inserted on them. 
The most familiar example is that of a fundamental string ending along a curve on the boundary of  AdS$_5$ within 
type IIB string theory, dual to the Wilson loop operator in $\N=4$ SYM theory \cite{Maldacena:1998im}. When the curve is a circle or infinite straight line, the Wilson loop is $1\ov  2$-BPS and it preserves a 1d conformal symmetry,\footnote{The full symmetry group is $OSp(4^*|4) \supset SL(2,\mathbb{R})\times SO(3)\times SO(5)$.} corresponding to a string worldsheet with AdS$_2$ induced geometry \cite{Berenstein:1998ij}. 
The  strong coupling expansion for the expectation value of the Wilson line 
and correlation functions of operators inserted along it 
is then controlled by the  fluctuations  \cite{Drukker:2000ep} of the fundamental superstring near the static configuration 
(see, e.g., \cite{Cooke:2017qgm,Giombi:2017cqn} and refs. therein).

One can generalize this to other branes in different AdS background ending along different dimensional 
submanifolds on the boundary (for example, D3-brane and D5-branes probes in AdS$_5$ describing line operators, surfaces 
and domain walls, see, e.g.,  \cite{Constable:2002xt,Drukker:2008wr}). In the most symmetric cases these branes have the world-volume metric of 
AdS$_{p+1}\times S^k$ with appropriate $p$ and $k$.

In this paper we study the simplest such example within M-theory: an M2-brane  probe ending along a surface on the boundary of AdS$_7$. The most symmetric configuration, which preserves 
half the supersymmetries of the bulk theory, is when the
3d world-volume of the M2-brane ends on a plane (or sphere) at the boundary. 
As the M-theory in the \adsss vacuum  is a dual description of the (2,0)  conformal theory,  
this configuration should be representing a supersymmetric surface defect operator in this 6d CFT
(for a recent discussion   and refs.  see   \cite{Drukker:2020dcz}). 
Our aim is to study this system beyond the classical brane limit by calculating its one loop fluctuation 
determinant and performing the holographic computation of 4-point correlators of the simplest local 
insertions into the surface operator. This is the direct analog of the calculation of insertions into the 
Wilson loops, captured by string fluctuations \cite{Giombi:2017cqn}.

Let us recall the case of insertions into Wilson loops in ${\cal N}=4$ SYM. The $1\ov  2$-BPS 
Maldacena-Wilson line along the $x^1$ direction and coupling to the scalar $\Phi_6$ is
$W = \Tr{\cal P }e^{\int dx^1 \left(i A_1+\Phi_6\right)}$. We can insert any adjoint valued 
operators into the loop, the most natural being the remaining five scalars $\Phi_a$ and 
the combination of field strength and scalar $\DD_i={\mathbb F}_{ti}\equiv iF_{ti}+D_i \Phi_6$. 
The latter, known as the displacement operator, represents small geometric deformations of the line. 
In the defect CFT (dCFT), the scalars $\Phi_a$ have dimension one and the displacement dimension two. This 
translates in AdS to fluctuation modes of the AdS$_2$ worldsheet with 
$m^2=0$ and $m^2=2$. Their 4-point functions were studied in \cite{Giombi:2017cqn} by 
expanding the string action to quartic order and performing AdS$_2$ Witten-diagram 
calculations on the worldsheet. This allowed to deduce the spectrum of some of the 
operators appearing in their OPE, providing further details on the strong coupling Wilson loop dCFT. For instance, the scaling dimension of the singlet scalar insertion which is dual in AdS$_2$ to a two-particle ``bound state" of string fluctuations along $S^5$ was found to be
$\Delta = 2-\frac{5}{\sqrt{\lambda}}+\ldots$.  
Recently, this result was confirmed by integrability techniques in \cite{Grabner:2020nis}, which also obtained several more orders in the strong coupling expansion. 

For the case at hand, the (2,0) supersymmetric 6d CFT  describing  multiple M5-branes  may be thought of as a $SU(N)$ 
generalization of a  free (2,0) tensor multiplet containing the $B_{mn}$-field with self-dual strength $H_{mnl}$, 
5 real scalars $\Phi^I$ and 4 symplectic Majorana fermions.  In this abelian theory the  locally-supersymmetric   surface operator 
analogous to the Wilson loop operator of \cite{Maldacena:1998im}   may be defined as
\cite{Gustavsson:2004gj,Drukker:2020dcz}\foot{The introduction of  a 
surface operator  with coupling to $B$-field 
\ci{Ganor:1996nf,Maldacena:1998im,Berenstein:1998ij}  is natural  by analogy with strings ending on D3-branes case, i.e.
in the picture where 
the dynamics of M5-branes  is described in terms of M2-branes  \cite{Strominger:1995ac} 
ending on strings coupled to $B$-field.}
\be \la{313}
V = \exp \Big(  \int d^2 \vx \big[ i \, \half  \e^{\m\n}   \del_\m X^m  \del_\n  X^n\, B_{mn}(X)   + \sqrt {g(X)} \,  \Phi_5 (X)\big] \Big)
\to
\exp \Big(  \int d^2 \vx \big[ i \,  B_{12}(X)   +   \Phi_5 (X)\big] \Big)
\ , \ee
where  $\Phi_5$  is  one of the 5  scalars of the (2,0) tensor multiplet, $X^m(x)$ are the 6d coordinates describing the surface and 
we  specified to the case   when the defect is a plane in the $(1,2)$ directions.\foot{Due to conformal invariance 
one  can consider  the defect with with  either planar or spherical  ($S^2$) geometry.}
The surface operator breaks the $OSp(8^*|4)$ supersymmetry  of the  6d theory to 
$[OSp(4^*|2)]^2$
with the bosonic subgroup 
$SO(2,2)\times  SO(4)\times  SO(4)= [SO(2,1) \times SU(2) \times SU(2)]^2$. Here $SO(2,2)$ corresponds to the 2d conformal symmetry, one $SO(4)$ to rotations in the transverse directions to the surface, and the second $SO(4) \subset SO(5)$ is the remaining $R$-symmetry that rotates the four scalars that do not couple to the operator.
As natural in defect CFT, one can consider correlation functions of operators inserted on the defect surface: the   basic   short  
multiplet  includes four transverse scalars $\Phi_a$ ($a=1,\cdots,4$)  with dimension $\Delta =2$,
four displacement operators\foot{The displacement operator describes transverse deformations of the 
defect (see for instance \cite{Billo:2016cpy} for a general discussion). For a defect with co-dimension $6-p$, the displacement operator $\DD^i$  may be defined via
$
\partial_{\mu} T^{\mu i} = \delta^{(6-p)}(x_{\perp}) \DD^i
$
where $T $ is the  ``bulk" stress tensor (the stress tensor of the 6d CFT), with $\Delta=6$. 
For the surface defect  ($p=2$)  the dimension of the displacement $\DD^i (i=1,\cdots,4)$ is  then 
$\Delta=3$.  In general, for CFT$_d$ with a co-dimension $d-p$ defect, $\Delta(\DD^i) = p+1$.} 
$\DD_i= {\mathbb H}_{12i}\equiv iH_{12i}+\del_i \Phi_5$  $(i=1,\cdots,4)$  with $\Delta=3$
and eight fermions with $\Delta= 5/2$. 

In the dual  description this $1\ov  2$-BPS surface operator is represented by 
a probe M2-brane with worldvolume ending on a plane at the $\mathbb R^6$ boundary, 
stretched along  $z$ of \adss  and  localized at a point in  $S^4$. 
The M2-brane probe is described by a $\kappa$-symmetric generalization of the Dirac-Nambu action 
(see, e.g., \cite{Bergshoeff:1987cm,deWit:1998yu}).
The induced 3-geometry in the static gauge  is  then \adst and      
as in \cite{Forste:1999yj,Drukker:2000ep} one finds that  the transverse   fluctuations of the M2-brane surface are 
represented by:  4~scalars $y^a$ ($S^4$ fluctuations)  with $m^2=0$, 
4~scalars $\xx^i$   (AdS$_7$  fluctuations transverse to the 3-surface)  with $m^2=3$
and 8 fermions  with $m^2={9\ov 4}$. The correlators of these  ``transverse" 
membrane fluctuations (and more generally their composites) should then define a 2d dCFT associated to the surface defect. 
Via  AdS$_3$/CFT$_2$ correspondence the dual boundary operators 
should have dimensions $\Delta= 2, 3$ and~${5\ov 2}$  matching those of the scalars, displacement operator  and fermions on the 
defect.\foot{In general,  in AdS$_{p+1}$/CFT$_p$ case   we have 
$\Delta(\Delta-p) = m^2$  for scalars  and $\Delta= m + p/2 $ for the fermions. In the string (Wilson loop) case $p=1$, while here $p=2$.} 

Below we compute the  correlators of the bosonic fluctuations $X^I= (\xx^i, y^a)$ as defined by the 
M2-brane action  in the inverse effective membrane tension  $\TT= a^3 T_2 = {2\ov \pi} N$ expansion ($a$ is the radius of AdS$_7$). 
They  should define  the large $N$ limit of the corresponding 6d correlators of the operators $\OO_I= ( {\mathbb H}_{12i}, \Phi_a)$ 
inserted on the planar ($\vx=(x^1, x^2)$) defect 
\be\la{01} 
\llangle \OO(\vx_1)  \cdots \OO(\vx_n)  \rrangle  ={ \langle   X(\vx_1) \cdots X(\vx_n) \rangle}_{_{\rm AdS_3}}  \ . \ee
A novel feature of the present M2-brane in \adss$\times  S^4$   case compared to the string in 
AdS$_5 \times S^5$ case in \cite{Giombi:2017cqn}
is the presence of the WZ term in the action that contributes non-trivially  to the 4-point correlator of the scalars $y^a$. 
This term 
$\sim \TT \int_4  \ep^{ABCDE }  Y_A dY_B \wedge  dY_C  \wedge  d Y_D \wedge  d Y_E \to 
{ iN\ov 32 \pi}  \int d^3 x\,  \ep^{\m\n\l} e_{abcd}\, 
y^a \del _\m y^b   \del _\n y^c \del _\l  y^d  +  O(y^5) $  
originates from the coupling of the M2-brane to  the potential $C_3 $ of the  magnetic 4-form flux  of the \adsss
background \cite{Freund:1980xh}.\foot{A similar term is present, e.g., in the D3-brane probe action in \adssp \cite{Tseytlin:1999tp}.}
Being intimately related to the underlying supersymmetry, the  contribution of this term is important for the resulting 4-point 
function satisfying the constraints imposed by the residual superconformal symmetry.
In contrast to the Wilson loop in $\N=4$ SYM case where one can also directly  compute  a weak-coupling 
limit of the corresponding correlators on the gauge theory side, it is  not clear how to do this in  the (2,0) 6d theory 
that currently lacks an intrinsic definition.\foot{One can still mimic such computation 
by starting with the abelian 6d tensor multiplet theory  and consider  correlators  of the fields with the defect \rf{313}. 
In particular,  ref. \ci{Gustavsson:2004dm}    computed
the 2-point function of the displacement operator  by considering  the second order
in  the  ``wavy surface" approximation. Its form  
$\langle  \DD^i(\vx_1)\DD^j(\vx_2)\rangle  \sim    {1  \ov  |\vx_{12}|^6}$ is  dictated  by the associated dimension  $\Delta=3$.} It would be interesting to make contact with the results of this paper by bootstrap methods, as was done in \cite{Liendo:2018ukf} for the case of the Wilson line dCFT.
\foot{Among possible generalizations one may consider a BPS configuration of a M5 brane probe 
intersecting M5 branes over a line and wrapped on $S^3\subset S^4$  
so that the resulting  M5 brane 
world volume  geometry is AdS$_3 \times S^3$ (cf. \ci{Berman:2001fs,Mori:2014tca,
DHoker:2008rje,Lunin:2007ab,Chen:2007ir}). This should correspond to the case when the surface defect 
is in a large representation of $SU(N)$. 
}

The contents of this paper are as follows. Our starting point in
Section~\ref{sec:action} is the expansion of the  M2-brane action in \adsss
near the minimal  3-surface ending on a 2-plane or a 2-sphere at the  boundary (and localized at a point in $S^4$). 
The    value  of the  classical M2-brane action  on this surface  is proportional to the volume of AdS$_3$. In the case of spherical boundary, the volume of \adst is logarithmically divergent with the IR cutoff R.
This is  in contrast to the string in the AdS$_5$ case, where the classical value of the string action proportional to the 
volume of AdS$_2$  is finite (after subtraction), and matches the strong-coupling limit of the expectation value of the circular Wilson loop. 
Here instead the coefficient  of $\log\RR$  term may be  interpreted as one of the conformal anomaly coefficients
in the  defect CFT.\foot{Similar logarithmic UV divergence appears in the log of  expectation value of 
the   surface operator \rf{313} in the abelian (2,0)  theory   \cite{Gustavsson:2004gj,Drukker:2020dcz}.
For the dual M2-probe discussion see also  \ci{Rodgers:2018mvq}.} 

In Section~\ref{sec:1-loop}  we compute the 1-loop correction to the logarithm of the partition function of the M2-brane 
ending on a spherical surface. This gives a correction of order $N^0$ to the leading result coming from the classical action of the surface, which is of order $ \TT = {2\ov \pi} N$. In a choice of normalization that will be explained below, we find for the anomaly coefficient of the spherical surface 
$b= 12 N-9  + \OO(N^{-1})$. These first two terms match the prediction $b= 3 (N-1) (4 + N^{-1})$
following from  \ci{Estes:2018tnu,Jensen:2018rxu}.
In Section~\ref{sec:neumann}  we also comment on the holographic description of the 
non-supersymmetric surface defect operator which does not couple to the scalar fields,   following the analogy with 
the standard Wilson loop  case  in \ci{Alday:2007he,Polchinski:2011im,Beccaria:2017rbe,Beccaria:2019dws}. 
In this case the  M2-brane surface should be delocalized in $S^4$, i.e. the  scalars  $y^a$ should satisfy the Neumann   boundary condition.
Adding a boundary perturbation to  the M2-brane action leads to a 2d RG flow   between the UV (non-supersymmetric) and IR (supersymmetric)   fixed points  with the resulting  values of the boundary 
conformal anomaly coefficients consistent with the b-theorem for 2d defects \cite{Jensen:2015swa,Casini:2018nym,Kobayashi:2018lil}.

In Section~\ref{sec:4pt} we compute the 4-point correlation functions for the scalar  fluctuations  $y^a$ and $\xx^i$ 
near the BPS  surface, in the leading tree-level approximation. We find  the expressions  following from the Dirac-Nambu part of the action 
for the general dimension $p$ of the brane, with the $p=1$ case reproducing the string-theory results of  \cite{Giombi:2017cqn}. We observe that the $\langle yyyy\rangle$ 4-point function satisfies simple superconformal Ward identities that, as turns out,
 essentially determine its form. We also discuss the Mellin representation for the resulting  AdS$_3$ correlators. 
In Section~\ref{sec:OPE} we perform an OPE  analysis of the correlator $\langle yyyy \rangle$  extracting the leading $1/N$ terms in the anomalous 
dimensions of composite $y \del^n y$ operators appearing in different channels.  

In Section~\ref{sec:super} we  discuss constraints on correlators imposed by a  residual superconformal symmetry. 
In Section~\ref{sec:twisted}   we follow the analogy with the SYM case \cite{Drukker:2009sf}
and  consider a special  twisted  combination ${\cal Y}=\tt^a(\vx) \, y^a(\vx)$  of scalar operator $y^a$   whose correlators are constrained by residual supersymmetry. 
The  4-point  correlator  of the twisted fields  has a very simple form  given in Section~\ref{sec:twisted4pt} 
and  surprisingly  has a very similar  structure to that of  
the strong-coupling limit of the reduced correlator of $1\ov  2$-BPS  scalar operators in the ${\cal N}=4$ SYM theory. 
In Section~\ref{sec:ward}   we  show that the form  of the $\langle yyyy \rangle$   correlator  found in Section~\ref{sec:yyyy} is essentially constrained  by the superconformal  symmetry and  crossing up to an overall constant factor.
 A  similar observation in the case  of the scalar   correlators on the  BPS Wilson loop in $\N=4$ SYM  is  made  in 
 Appendix~\ref{sec:WL}.

\section{Membrane  action in AdS$_7 \times$ S$^4$}
\label{sec:action}

We are interested in studying the fluctuations of an M2-brane in the  AdS$_7 \times S^4$ background 
which is the near horizon geometry of $N$ M5-branes
\begin{align}
 \la{45}
& ds^2 = a^2\big[ ds^2_{_{{\rm AdS}_7}}+\rr^2 ds^2_{_{ S^4}}\big]
\ ,\qquad a^3 = 8 \pi N \ell_p^3 \ , \quad  \rr=\frac12 \ , \\
& F_4= \pi^2 a^3 \Omega_4 \ , \qquad   \int_{S^4}  \Omega_4=1 \ , \qquad
\vol(S^4) = \frac{8 \pi^2 }{3} \ . \la{310}
\end{align}
Here $a$ is the radius of  AdS$_7$, and $ds_{_{{\rm AdS}_7}}$ and $ ds_{_{ S^4}}$ are the line elements 
on unit radius AdS$_7$ and $S^4$. \  $\Omega_4$ in \rf{310} denotes the  normalized 
volume form of $S^4$, 
 $\ell_p$ is defined via  $2\kappa_{11}^2 = (2\pi)^8 \ell_p^9$
 and $F_4 = d C_3$.    
We  use Euclidean signature throughout. 

The (bosonic part of) the $\kappa$-symmetric  action   of an M2-brane probe  
\cite{Bergshoeff:1987cm} 
contains two terms: the standard Dirac-Nambu type term $S_1$  (the volume of the  3-surface in  induced metric) 
and a WZ-type term $S_2$ of coupling to the 3-form $C_3$ 
\begin{align}
\la{31}
S= S_1  + S_2 \,: \qquad 
S_1&= T_2 \int d^3 x \, \sqrt {\det h_{\m\n} } \ , 
\qquad  h_{\m\n} = \del_\m X^M \del_\n X^N G_{MN} (X) \ , 
\\
S_2 &=-  i T_2 \int d^3 x \, \tfrac{1}{3!} \ep^{\m\n\l} C_{MNK}(X)\,  \del_\m X^M \del_\n X^N \del_\l X^K  \ ,  \la{32} 
\end{align}
where the fundamental M2-brane tension $T_2$ is 
\be\la{4450}
{ T}_2= (2 \pi)^{2/3}   (2\kappa^2_{11})^{-1 /3}  = {1\ov (2\pi)^2 \ell_p^3} \ .
\ee
The world-volume on an M2  brane
ending on  the plane (or sphere)  at the 6-boundary of AdS$_7$ has the 
classical solution with AdS$_3$ induced metric.\foot{
The existence of such static  M2 brane   solution is related to the fact 
that  M2  brane   intersecting with a stack of M5 branes over a plane  is a $1\ov  2$-BPS configuration.
This 
can be easily seen, e.g.,   from the  absence of force on a static M2 brane in this case   \cite{Tseytlin:1996hi} ($C_3$ here is purely magnetic).}
In the case of the planar surface, this AdS$_3$ subspace is spanned by 
the plane and the $z$ coordinate in \eqref{45}.

The effective membrane tension (i.e. the analog of the familiar  fundamental string tension 
${\rm T}_1={a^2\ov 2 \pi \a'}= {\sqrt \lambda \ov 2 \pi}$ in the AdS$_5 \times S^5$ case)
is then  given by  the product of the fundamental M2-brane tension $T_2$ 
and the cube of the AdS radius, i.e.
(see, e.g., \ci{Klebanov:1996un,Tseytlin:2000sf})\foot{In the notation of 
\cite{Klebanov:1996un,Tseytlin:2000sf}
$ 2 \pi \ell_p^3 = \ell_{11}^3$.} 
\beg \la{2.7}
\TT=  a^3   T_2 = {2\ov \pi} N  \ . \ee

 We proceed now to expand the  probe brane action about this 
classical solution.
For generality, let us 
consider a $p$-brane in AdS$_{d+1}$  with  world volume  ending  along a $p$-plane  at the boundary 
and also stretched along $z$ of AdS$_{d+1}$. Following \cite{Giombi:2017cqn}  where the case of $p=1$ and $d=4$  was discussed, 
let us choose the following ``AdS$_{p+1}$-adapted"  parametrization of AdS$_{d+1}$ (with radius 1) 
 \beg  \la{33} 
ds^2_{d+1}  =\frac{ (1+\fo  \xx^2)^2}{(1-\fo \xx^2)^2} ds^2_{p+1}  + \frac{d \xx^i d\xx^i}{(1-\fo \xx^2)^2} \ ,\qquad
ds^2_{p+1} =\frac{1}{\xxxx^2} (d\xxxx^2 + d\xx^v  d\xx^v)      \    , 
\eeg
where the indices of the  boundary coordinates  of  AdS$_{d+1}$  are split  into    $v=1, \cdots,p$ and   $ i=1, \cdots, d-p$.
The minimal   surface  ending on  a $p$-plane  at the boundary  is 
\beg \la{34}  \xx^v = x^v \ , \qquad  \xxxx= \xxx  \ ,   \qquad \xx^i=0\  ,  \qquad 
ds^2_{p+1} = \frac{1}{\xxx^2} (d\xxx^2  + dx^v  dx^v)\equiv  g_{\m\n}(x)d x^\m dx^\n\ , 
\eeg 
so that the corresponding induced metric is AdS$_{p+1}$. 
Choosing a static gauge  in the $p$-brane action in AdS$_{d+1} \times S^n$ we get for the $S_1$ part of  its action in \rf{31} 
\beg \la{35} 
S_1 =  {\rm T}_p  \int d^{p+1} 
x  \,  \sqrt{\det \bigg[    \frac{(1+\fo  \xx^2)^{2}}{(1-\fo \xx^2)^{2}}  g_{\m\n}   + \frac{\del_\m \xx^i \del_\n \xx^i  }{(1-\fo \xx^2)^2}
+  
{  \partial_\m y^a\partial_\n y^a \ov (1+\tfrac{1}{4\rr^2}   y^2)^2}  \bigg] } \equiv   {\rm T}_p \int d^{p+1}x   \sqrt{ g}\  L  \ , 
\eeg
where $y^a$   are coordinates of $S^n$  and $\rr$  is its radius in units of the radius of AdS$_{p+1}$
(which is absorbed into  the dimensionless effective tension $ {\rm T}_p $).

Expanding \rf{35}   in powers of the fluctuations  $\xx^i$ and $y^a$ we get  
\begin{align}   L &=L_2  + L_{4\xx}   + L_{2\xx,2y} + L_{4y}   + \ldots   \ , \qquad 
L_2=\tet  \frac{1}{2} \big[ g^{\m\n}\del_\m \xx^i \del_\n \xx^i  +  (p+1)\, \xx^i \xx^i\big] + \frac{1}{2}  g^{\m\n}\del_\m y^a \del_\n y^a\ ,  \la{36} 
\\[2pt]
L_{4\xx} &=\tet  \frac{1}{8} (g^{\m\n}\del_\m \xx^i \del_\n \xx^i )^2  
           - \frac{1}{4}  (g^{\m\n} \del_\m \xx^i \del_\n \xx^j) \; (g^{\rho\kappa} \del_\rho \xx^i \del_\kappa  \xx^j)
\nonumber     \\
       & \tet\quad + \frac{1}{4}  p\,  \xx^i \xx^i  \, g^{\m\n} \del_\m \xx^j \del_\n \xx^j + \frac{1}{8} (p+1)^2   \xx^i \xx^i\, \xx^j \xx^j \ , 
\label{37}            
\\[2pt]
L_{2\xx,2y}&=\tet  \frac{1}{4}  (g^{\m\n}\del_\m \xx^i \del_\n \xx^i )\,(g^{\rho\kappa} \del_\rho y^a \del_\kappa  y^a) 
    - \frac{1}{2}   (g^{\m\n} \del_\m \xx^i \del_\n y^a) \; (g^{\rho\kappa} \del_\rho \xx^i \del_\kappa  y^a) 
\no\\
&  \quad       
+  \tfrac{1}{4}   (p-1)   \xx^i \xx^i  \, g^{\m\n} \del_\m y^a \del_\n  y^a  \ , \la{38}
\\[2pt]
L_{4y} &= \tet
-\frac{1}{4\rr^2} y^b y^b\ g^{\m\n} \del_\m y^a \del_\n y^a
+\frac{1}{8}   (g^{\m\n}\del_\m y^a \del_\n y^a)^2
-\frac{1}{4}  (g^{\m\n} \del_\m y^a \del_\n y^b) \; (g^{\rho\kappa} \del_\rho y^a \del_\kappa  y^b)\ . \la{39}
\end{align}
The string in AdS$_5 \times S^5$ case   considered in  \cite{Giombi:2017cqn}  corresponds to $p=1,\ d=4,  \ n=5, \ \rr=1$ 
while in  the present  M2-brane in AdS$_7 \times S^4$ 
case we have $p=2, \  d=6, \ n=4, \ \rr=\ha $. 
In the latter case we get  4 transverse \adss  fluctuation fields  $\xx^i$ having  $m^2=3$ and 4 massless  $S^4$ fields  $y^a$
propagating in induced \adst geometry. 
One may  also  include the fermionic  terms  coming   from  the corresponding 
AdS$_7 \times S^4$   supermembrane  action as discussed in \cite{Forste:1999yj}  getting (after fixing $\kappa$-symmetry gauge) 
eight   3d  fermions with  $m= 3/2$.

To find the explicit expression for  the WZ term \rf{32}  in  the M2-brane   action   let  us note that the  
 normalized  volume form  $\Omega_4$ of a  unit-radius $S^4$  in \rf{310}
 may be  expressed  in  terms of a  unit 5-vector 
$Y^A$    as\foot{In general,  one has 
$\int_{S^d} \Omega_d = 1 \ , \ \Omega_d = { 1 \ov \vol(S^d)\,  d!} \ep_{d+1}  Y  ( \wedge dY)^d $, 
$\vol(S^d) = { 2 \pi^{d+1 \ov 2} \ov \Gamma(  {d+1 \ov 2}) }$. 
$\Omega_d$  appears in the expression for the  Hopf index of
the  map $S^d \to S^d$.
The associated topological current is 
$J^\l =  {1 \ov \vol(S^d)\,  d!}   \ep^{\l \mu_1\cdots \mu_d} \ep_{A B_1\cdots B_d}  Y^A  \del_{\m_1} Y^{B_1} \cdots \del_{\m_d} Y^{B_d} 
$, \  $ \int_{S^d}  J^0=N$=integer.}
\be \la{388}
\Omega_4 = {1 \ov 64 \pi^2}  \ep_{ABCDE} Y^A dY^B \wedge dY^C \wedge dY^D \wedge dY^E \ , \qquad \qquad 
Y^A Y^A =1\ , \ \ \ A=1,\cdots,5\ . 
\ee
Using the expression \rf{2.7} for the effective M2-brane tension 
the WZ term in \rf{32}  takes the form 
\be \la{399}
S_2=- iT_2  \int  C_3  =  - iT_2 \int  F_4   =
-    {i N  \ov  32 \pi }  \int d^4 x  \  \ep_{ABCDE}\, \ep^{\m\n\l\rho}\,   Y^A \del_\m Y^B \del_\n Y^C \del_\l Y^D \del_\rho Y^E \ . 
\ee
Like in the case of a similar WZ term in the D3-brane case \ci{Tseytlin:1999tp}
a manifestly $SO(5)$ invariant  form of  the 
WZ term is non-local---given by an  integral  over a  4-surface  that 
has the world-volume as its boundary. The  normalization of \rf{399} is checked by observing that if the 4-surface  is $S^4$ 
the integral in  \rf{399}   becomes $- 2\pi i N$, i.e. $e^{-S_2}= 1$.\foot{
Being topological this WZ term   should not be renormalized   and should be derivable 
as in  \ci{Tseytlin:1999tp,Intriligator:2000eq,Abanov:1999qz}
from the 1-loop
fermionic determinant  in the dual 6d theory in the presence of a  defect  represented by the surface operator.}

Setting 
\be \la{411}
Y^5=  \frac{1-\frac{1}{4\rr^2}   y^2}{ 1+\frac{1}{4\rr^2}   y^2}\ , \ \   \qquad 
Y^a=  \frac{ \frac{1}{\rr}y^a}{ 1+\frac{1}{4\rr^2}   y^2}\ ,\qquad \qquad  Y^AY^A=1 \ , 
\ee
where we rescaled $y^a$ by $\rr$ to conform with \rf{35}, 
we find that the expansion of \rf{399}  in powers of  $y^a$ starts with  the $y^4$ term 
(as  $\rr= \ha$   we have  $ 32 \pi \rr^4 = 2 \pi$) 
\be \la{40}
S_2 = -  { i N   \ov 2 \pi }   \int d^3 x \, \ep^{\m\n\l}\,  \ep_{abcd} \,  y^a \del_\m y^b  \del_\n y^c  \del_\l  y^d + \OO(y^5)   \ . 
\ee 
The  explicit normalization of the kinetic term for $y^a$ in \rf{35} is (using \rf{2.7})
\be \la{41}
S_1= {N\ov \pi} \int d^3 x \, \sqrt g\, g^{\m\n} \del_\m y^a \del_\n y^a + \ldots \ . \ee

\section{One-loop partition function: defect conformal anomaly}
\label{sec:1-loop}

In this section we calculate the fluctuation determinants about the AdS$_3$ classical 
M2-brane solution. 
The more complicated problem of deriving the 1-loop quadratic fluctuation for 2 parallel 
planes was discussed in \ci{Forste:1999yj}. The fluctuation spectrum presented in 
the preceding section indeed matches their spectrum in the limit of large separation.
The discussion is parallel to the one in the string case in \ci{Forste:1999qn,Drukker:2000ep}. 

To recall, our spectrum has  4 bosons with $m^2=3$ plus  4 bosons with $m=0$     and
8 fermions  with  $m= 3/2$.%
\foot{%
The  bosonic  and  fermionic operator are   essentially universal  for the 
straight line or two parallel
line configurations:   what changes is just the induced geometry.
The same   was in the case of a  string in AdS$_5 \times S^5$: there one had \ci{Forste:1999qn,Drukker:2000ep}:
2 bosons   with $m^2 = 2/a^2$;
1 boson with   $m^2 = 4/a^2  + R^{(2)}$;
5  massless  $S^5$    bosons; 
8 fermions  with $m= {1/ a}$  or   squared operator  $- \nabla^2 +\fo  R^{(2)} +  {1/a^2}$. 
In the  straight-line or circular line surface  the induced geometry was AdS$_2$ so 
$R^{(2)} = -2/a^2$. 
It is remarkable that the structure of the partition function
in the string and M2 cases is very similar. This has to do,  in
particular, with the universal form of the Nambu-type term in the p-brane action 
and also the fact that in a natural $\kappa$-symmetry gauge  the 
fermionic kinetic term comes from the supergravity  covariant derivative  projected to the world volume 
that  contains  the $F$-flux term 
that gets contribution  from the sphere  magnetic part that is not
sensitive to  the details of surface in  the AdS   space.}
The resulting partition function $Z$ is then  given by 
\beg \la{4.1}
\F_{\rm 1-loop}= - \log Z= \half \big[4  \log \det   ( - \nabla^2 + 3)   + 4  \log \det  ( - \nabla^2 ) -  8 \log \det \Delta_{1/2} \big]  \ . 
\eeg
To evaluate the determinants we may follow the approach used  in the AdS$_5 \times S^5$   string case  in 
\ci{Drukker:2000ep,Buchbinder:2014nia}, i.e.  use the results for heat kernels of operators in AdS space 
from \cite{Camporesi:1994ga}.

Cubic UV divergence cancels out   due to the equal number of bosons  and fermions. 
The linear divergence of $\log \det(-\nabla^2 + X)$  in 3d is 
proportional to $b_2= \tr ( {1\ov 6} R^{(3)} - X)$   so here   $b_{2\, \rm tot}= -6= R^{(3)}$ 
and thus, as in string case, is proportional to the Euler number (assuming boundary terms are taken into account, cf. footnote 30 in
\ci{Drukker:2000ep}). In any case, such divergences are absent in an analytic  regularization like $\zeta$-function one and may be ignored. 

There   is no  bulk  logarithmic   UV divergence in 3d   but the resulting  UV finite   $\F$ contains   a  logarithmic  IR divergence.
It can be found   using  the results in Section~3 of~\cite{Giombi:2013fka} (see also \cite{Giombi:2016pvg}), 
which studied the determinant of higher spin theory on $AdS_3$.  
For a scalar  of mass $m$ (with $m^2=\Delta(\Delta-2)$), the contribution to $\F$ is%
\foot{The general formulas for the AdS$_{d+1}$ 
spectral density for bosonic totally symmetric rank-$s$ and fermionic in the $[s,1/2,\cdots1/2]$
representation in general boundary dimension $d$ (see \cite{Camporesi:1994ga})
are presented  in \cite{Giombi:2016pvg} (see eqs. (3.20)  and (3.22)).
For $d=2$, they happen to coincide as a function of $s$.} 
\beg \la{4.2} 
\F_0^{ (\Delta)} =\te  {1\ov 2} \log \det ( -\nabla^2 + m^2) = - { 1\ov 12 \pi } (\Delta-1)^3  \,  {\rm \vol}({\rm AdS}_3) \ . 
\eeg 
For spin 1/2 fermion   with  $\Delta = p/2 + m = 1 + m$  we get 
\beg \la{4.3} 
\F_{1/2} ^{ (\Delta)} =\te   {1\ov 2} \log \det ( -\nabla^2 + \tfrac{1}{4}R + m^2)  = -
{ 1\ov 12 \pi } (\Delta-1)\big[\Delta(\Delta-2)+{1\over 4}\big]    \,   {\rm \vol}({\rm AdS}_3) \ . 
\eeg 
Introducing  $\RR$ as   the \adst IR cutoff  regularizing the \adst volume (e.g.  the radius of the boundary $S^2$)
we have  (for the unit-radius AdS$_3$)\foot{In general, 
 the regularized volume of  global AdS$_{p+1}$ space with  $S^p$ as its  boundary for   even $p$ is, discarding power-law divergences
  (see, e.g., \ci{Diaz:2007an}): \ \ 
$ \vol({\rm AdS}_{p+1})=
\frac{2(-\pi)^{p/2}}{\Gamma\big(1+\frac{p}{2}\big)}\log \RR\,. $ }
\be \la{42}
{\rm \vol}({\rm AdS}_3) = - 2 \pi   \log \RR  \ . \ee
As a result,  we get for \rf{4.1} 
\beg\la{4.4} 
\F_{\rm 1-loop} = 4 \F_0^{ (\Delta=3)} + 4 \F_0^{(\Delta=2) } - 8 \F_{1/2}^{(\Delta=5/2)} = 3 \log \RR \ . 
\eeg 

\subsection{Interpretation of the result}
Equation \eqref{4.4} is the 1-loop correction to the 
tree level contribution  given by the value of the M2-brane action which is  just the  M2-brane tension \rf{2.7}
times 
the regularized  volume of the induced \adst metric  (cf. \cite{Berenstein:1998ij}) 
\beg \la{4.55}
\F_{\rm tree} = \TT {\rm \vol}({\rm AdS}_3) = -2\pi \TT  \log \RR  \ . 
\eeg 
The coefficient of $\log \RR$  in  equation \rf{4.55} and in  \eqref{4.4} 
has the interpretation of a conformal anomaly coefficient in the defect CFT. 
Surface operators have three anomaly coefficients,%
\footnote{A fourth can be defined for nontrivial coupling to the scalar fields, see \cite{Drukker:2020dcz}. 
By adjusting the scalar coupling, 
the total anomaly of some BPS surface operators, different from the sphere studied here, 
vanishes \cite{Mezei:2018url}.}
each multiplying a particular conformally invariant integral on the surface related to its topology, 
extrinsic curvature and background Weyl tensor (see \cite{Schwimmer:2008yh} for details). 
Since our calculation is focused on the single surface geometry of the sphere 
(the plane has trivial anomaly), our result captures one particular combination of the anomaly coefficients 
which we denote as $b$. Thus $\F$ can be expressed as
\beg \la{4.6}
\F \equiv - \tfrac{1}{3} b  \, \log \RR \ , \qquad \qquad 
b = -{3} (- 2\pi \TT + 3 )  + \ldots = {12 }  N -  9 + \ldots\ , \eeg 
where  dots stand  for possible higher-loop $1/\TT \sim 1/N $ terms.

The leading order at large $N$ of the other two anomaly coefficients were calculated holographically by 
Graham and Witten \cite{graham:1999pm} by considering M2-branes ending on arbitrary surfaces. They can also be 
inferred in other ways: the coefficient related to extrinsic curvature is proportional to the normalization of the 
displacement operator, as mentioned in Section~\ref{sec:4pt} below eq.~\rf{666}. The remaining one was 
conjectured to also be fixed by the same normalization constant in theories with enough 
supersymmetry \cite{Bianchi:2019sxz} (based on 
\cite{Lewkowycz:2013laa, bianchi:2018zpb}) as is indeed verified in
 \cite{Drukker:2020atp}.

Going back to our expression for the coefficient $b$ in \eqref{4.6},  we observe that  it happens to be 
  consistent with the result for the  corresponding anomaly 
coefficient found in \cite{Estes:2018tnu,Jensen:2018rxu} from the entanglement entropy for the 
``bubbling"  M5-M2 geometry  with M2-branes  corresponding to a   $1\ov  2$-BPS  surface defect 
operator in (2,0)  theory   in a  $su(N)$ representation  with the  Young tableau  with a large number  of boxes.\foot{An exact expression for another anomaly coefficient is derived in \cite{Chalabi:2020iie} from
the computation of the associated superconformal index.}
In the notation of \cite{Estes:2018tnu}   we have 
\be \la{4.77}
b= 24(\rho, \lambda) +3 (\lambda, \lambda)   \ , \ee
where $\rho$ is the Weyl vector of  $su(N)$   and $\lambda$ is the  highest weight of a  particular $su(N)$ 
representation.
If we formally   assume  that  this expression  should be valid not  just for large representations but also for the ones
 with finite number of boxes  then in 
 the present case of a single M2-brane  corresponding to the surface operator in the fundamental representation one   finds $(\rho, \lambda)= {N-1\ov 2} , \  (\lambda, \lambda) = {N-1\ov N}$ and thus 
\be \la{4.8}
b  = 3 (N-1) ( 4 + {N^{-1}} ) =   12 N - 9  - {3N^{-1}}    \  . \ee
Remarkably, this is in agreement with \rf{4.6}  and 
suggests  that the perturbative expansion in \rf{4.6} 
may terminate after the 2-loop ${1\ov N}$ term.\foot{If one assumes that the series in \rf{4.6} terminates at $1/N$ order then 
the coefficient of this term can be of course fixed  by requiring that the full expression should vanish for $N=1$.}
It would  be very interesting to compute this term directly from the 2-loop supermembrane
 Witten diagrams in AdS$_3$.\foot{For comparison, let us recall  the expressions for 
the conformal anomaly 
coefficients of the ``bulk" theory---the   $(2,0)$    theory describing $N$ coincident 
M5 branes (see, e.g., \cite{Beccaria:2015ypa}  for a review): 
$a=  -\frac{1}{4 \times 288}\, (16 N^{3}- 9\,N- 7)=  -\frac{1}{288}\, (N-1)\big[( 2N+1)^2 + { 3\ov 4}\big] $, \ 
$c = -\frac{1}{288}\, (4 N^{3}-3 N -1)=   -\frac{1}{288}\, (N-1) ( 2N+1)^2$.
The leading  $N^3$ terms   follow    \ci{Henningson:1998gx}
from the classical supergravity 
action,  
the   order $N$ terms   originate from the  $R^4$   corrections in 11d action  \ci{Tseytlin:2000sf}
and order $N^0$ terms  are from  the 1-loop 11d  supergravity 
corrections \ci{Beccaria:2014qea,Mansfield:2003bg}. 
The exact expressions follow also  from  non-perturbative approaches   based on 
supersymmetry  constraints   \cite{Beem:2014kka,Ohmori:2014kda}.
}

\iffa 
Going back to our expression for the coefficient $b$ in \eqref{4.6},  we observe that it
is  consistent with the result for the  corresponding anomaly 
coefficient found in \cite{Estes:2018tnu,Jensen:2018rxu} from the entanglement entropy for the 
``bubbling"  M5-M2 geometry (with M2-branes  corresponding to a   $1\ov  2$-BPS  surface defect 
operator in (2,0)  theory   in a  $su(N)$ representation  with the  Young tableau  with a large number  of boxes).%
\foot{An exact expression for another anomaly coefficient is derived in \cite{Chalabi:2020iie} from
the computation of the associated superconformal index.}
In the notation of \cite{Estes:2018tnu}   we have 
\be \la{4.77}
b= 24(\rho, \lambda) +3 (\lambda, \lambda)   \ , \ee
where $\rho$ is the Weyl vector of  $su(N)$   and $\lambda$ is the  highest weight of a  particular $su(N)$ 
representation.  In the present case of a single M2-brane  corresponding to the surface operator in the fundamental representation one   finds $(\rho, \lambda)= {N-1\ov 2} , \  (\lambda, \lambda) = {N-1\ov N}$ and thus 
\be \la{4.8}
b  = 3 (N-1) ( 4 + {N^{-1}} ) =   12 N - 9  - {3N^{-1}}    \  . \ee
This implies  that the perturbative expansion expansion in \rf{4.6} 
should terminate after the 2-loop ${1\ov N}$ term.\foot{If one assumes that the series in \rf{4.6} terminates at $1/N$ order then 
the coefficient of this term can be of course fixed  by requiring that the full expression should vanish for $N=1$.}
It would  be very interesting to compute this term directly from the 2-loop supermembrane
 Witten diagrams in AdS$_3$.\foot{For comparison, let us recall  the expressions for 
the conformal anomaly 
coefficients of the ``bulk" theory---the   $(2,0)$    theory describing $N$ coincident 
M5 branes (see, e.g., \cite{Beccaria:2015ypa}  for a review): 
$a=  -\frac{1}{4 \times 288}\, (16 N^{3}- 9\,N- 7)=  -\frac{1}{288}\, (N-1)\big[( 2N+1)^2 + { 3\ov 4}\big] $, \ 
$c = -\frac{1}{288}\, (4 N^{3}-3 N -1)=   -\frac{1}{288}\, (N-1) ( 2N+1)^2$.
The leading  $N^3$ terms   follow    \ci{Henningson:1998gx}
from the classical supergravity 
action,  
the   order $N$ terms   originate from the  $R^4$   corrections in 11d action  \ci{Tseytlin:2000sf}
and order $N^0$ terms  are from  the 1-loop 11d  supergravity 
corrections \ci{Beccaria:2014qea,Mansfield:2003bg}. 
The exact expressions follow also  from  non-perturbative approaches   based on 
supersymmetry  constraints   \cite{Beem:2014kka,Ohmori:2014kda}.
}

\fi

\subsection{Non-supersymmetric  surface defect and 2d RG flow}
\label{sec:neumann}

The    above  discussion  applied to the  case of the dual  description of the $1\ov  2$-BPS   surface  operator  which 
(at least in the abelian case)  should   be represented  by an analog of the Wilson-Maldacena  exponent \rf{313}
with one of the 5  scalars of the (2,0) tensor multiplet  coupled to the induced metric 
as in \cite{Gustavsson:2004gj,Drukker:2020dcz}.
This  
breaks $SO(5) $ R-symmetry to $SO(4)$   and corresponds, in the M2-theory in AdS$_7 \times S^4$, 
to an expansion near a point of $S^4$  with 4  $S^4$  massless scalars  subject to the Dirichlet  boundary condition  in AdS$_3$. 

By analogy with the Wilson loop case 
\ci{Alday:2007he,Polchinski:2011im,Beccaria:2017rbe,Beccaria:2019dws} 
we may also consider the  dual description of 
non-supersymmetric surface operator  without scalar  coupling \cite{Ganor:1996nf} preserving $SO(5)$  symmetry. 
In this  case  the classical M2-brane minimal   surface is the same AdS$_3$  but 
one is  to impose  the Neumann  boundary condition on $S^4$  fluctuations (and average over 
an expansion point in the sphere) to preserve the $SO(5)$   symmetry. 

As in the case of the surface for the circular Wilson loop in AdS$_5 \times S^5$  the 1-loop 
contributions of the  4  massive  AdS$_7$ scalars  and 8  fermions in \rf{4.1}  remain the same as in \rf{4.4} 
while to find the contribution of the 4 massless  $S^4$   fluctuations with alternative b.c.
(i.e.  with $\Delta=\Delta_- = 0 + \OO({1\ov N})$)  we may use, e.g.,  the  relation 
\ci{Diaz:2007an,Giombi:2013yva,Beccaria:2014jxa}
between the 
AdS$_{d+1}$ bulk  field  and  $S^d$ boundary  conformal  field partition functions:
$  Z^{(\Delta_-)}/Z^{(\Delta_+)} = Z_{\rm conf}$.  For a   scalar in  AdS$_{d+1}$   
one has   $\Delta_\pm= {d\ov 2} \pm \nu, \ \ \nu=\sqrt{{d^2\ov 4}+ m^2}$ so that 
the boundary  conformal (source) field with 
canonical dimension $\Delta_- = d - \Delta_{+}$
has the  kinetic term $\int d^d x \, \varphi (-\nabla^2)^\nu  \varphi$.

In the present case  $d=2, \ \Delta_+=2, \ \Delta_-=0, \ \nu=1$
so that the induced boundary CFT$_2$ 
has  the standard  kinetic operator $ -\nabla^2 $    on $S^2$, i.e.  the difference of the scalar free energies is  
\be 
\la{4.88}
\F_0^{(\Delta_-)} - \F_0^{(\Delta_+)} = \te  { 1 \ov 2}  \log {\det}' ( - \nabla^2)= - {1\ov 3} \log \RR  + \ldots \ . 
\ee 
where $\RR$ plays the role of a UV cutoff in 2d. 
The  positivity of the difference is in  agreement 
with the expected  ``defect $b$-theorem" \cite{Jensen:2015swa,Casini:2018nym},  
viewing, by analogy with the circular Wilson loop case \ci{Polchinski:2011im,Beccaria:2017rbe},
the non-BPS surface operator as the UV limit 
deformed by the relevant operator $Y_5 \sim \Phi_5$  to flow to the  BPS  surface operator, i.e. 
$  b_{_{\rm UV}}-b_{_{\rm IR}} = +1 > 0 $.  

As a result, taking into account the multiplicity 4 of $S^4$ 
scalars in \rf{4.1}, we conclude that   in  the  non-supersymmetric  (non-BPS)  defect 
case we should get instead of  the  $b=b_{\rm susy}$ in \rf{4.6}\foot{Compared to the non-supersymmetric 
circular Wilson loop case in \ci{Beccaria:2017rbe}   here  the $S^4$  zero mode  contribution $\sim 4 \log N$ 
appears only in the finite part of $F$, i.e. is not relevant for  the  1-loop conformal anomaly.}
\beg \la{4.66}
b_{\rm non-susy} = b_{\rm susy}  + 4 =  {12 }  N -5 + \ldots\ . \eeg 
One  may attempt to understand the RG flow  between the  non-supersymmetric and supersymmetric 
cases   by using  the same approach   as in  the string theory description of the circular Wilson loop \cite{Polchinski:2011im,Beccaria:2017rbe}. Starting with the (super)membrane   action\foot{One may  expect   that the 
standard first-derivative  supermembrane  action   in AdS$_7 \times S^4$  is  not renormalized (i.e. tension is not renormalized):
it  contains  fermionic and bosonic WZ  terms that  can not be renormalized, 
and they  are related by $\kappa$-symmetry  to the rest of the terms (this is analogous, e.g., 
to non-renormalization of 11d supergravity action).  
Loop corrections may induce higher-derivative terms but presumably they should  not be relevant for the discussion below.}
\rf{31},\rf{32}
one may perturb it  by a 2d boundary term (here we concentrate only on the $S^4$ fluctuations, see 
\rf{411},\rf{41})
\be \la{51}
S_1=\ha  \TT  \int d^3 x \,  (\sqrt g\, g^{\m\n} \del_\m y^a \del_\n y^a + \ldots)    -  \kk\,  \TT \int d^2 \vx \,\sqrt {g_{_2}}\  Y_5 \ , \ \ \ \qquad
Y_5 = \sqrt{1- Y_a Y_a}= 1 - 2 y_a y_a  + \ldots   
\ .  \ee
Here  $\sqrt {g_{_2}} = {1\ov z^2}\big|_{z\to 0}$  and $\kk$ is a new coupling  which  will run  between the UV and IR fixed points  (see Section~4.2 in \cite{Beccaria:2017rbe}  for details). 
$\kk Y_5$   term should  correspond to  a  similar scalar  $\Phi_5$ term in the exponent in the  surface 
operator (cf. \rf{313}) with coefficient running between 0 and 1.  
The variation  of \rf{51}     implies  that     $y^a$    should 
satisfy  the  massless   wave  equation in \adst  with  the metric $ds^2 = {1\ov \xxx^2} ( d \xxx^2 + d \vx d \vx)$  subject to the  Robin  boundary condition ($\del_n = n^\mu \del_\mu, \ \ n^\mu= ( -z, 0, 0)$) 
\be \la{552}
\big(\del_n - 4  \kk\big) y^a  \Big|_{\xxx\to 0} =0\  ,  \qquad \  \ \ \ \  \del_n = - z \del_z \ .     
\ee 
The parameter $ 0 \leq  \kk \leq \infty$ thus   interpolates between the Neumann ($\kk=0$)
and Dirichlet  ($\kk\to \infty $) boundary  conditions corresponding to 
$y^a=  \xxx^{\Delta_-}  v^a+ \OO(\xxx^2) =  v^a  + \OO(\xxx^2) $
and $y^a= \xxx^{\Delta_+}  u^a  + \OO(\xxx^2) = \xxx^2\, u^a   + \OO(\xxx^2) $ respectively. 
$\k$ will be running with UV   scale $\L$  of  the 3d theory.   Integrating $y^a$ out we get at leading 1-loop 
order the following    boundary  divergence  (ignoring power divergent terms)
\begin{align}  \la{553}
& \G_\infty= 4 \times  \half \log \det ( -\nabla^2  )\Big|_{\infty}  =  -  4 A_3 \log\L   + \ldots 
\ ,  \qquad
A_3 =    \tfrac{2}{  \pi} \int d^2 \vx \,\sqrt {g_{_2}}\   \ka^2 + \ldots   \ , 
\end{align}
where  $A_3$  is the  relevant Seeley  coefficient  (see, e.g., eq.~(5.32) in 
\cite{Vassilevich:2003xt}  where the Robin parameter $S$  is equal to $4\kk$ here). 
Adding \rf{553}  to  the bare action \rf{51}  and taking into account  that $\kk$ has  canonical dimension 2  we conclude  that the renormalized $\kk$ should  run according to (cf. \cite{Beccaria:2017rbe})
\be \la{554}
\beta_\kk = \mu {d \kk \ov d \mu} =  - 2 \kk  -     {8\ov \pi \TT}\, \kk^2  + \ldots 
= - 2 \kk   -     {4\ov N }\, \kk^2  + \ldots  \ . \ee
Here we used \rf{2.7}. 
As a result, we get  the UV fixed point at  $\kk=0$ 
and also a  possible  IR  fixed point at  $\kk= - {1\ov 2}N$. Assuming the latter can be trusted in the large $N$ expansion  it should represent the  Dirichlet limit of the Robin b.c. \rf{552}. 
Since the  derivative of the $\b$-function gives anomalous dimension at a
fixed point  the total   dimension ($2 +  \beta'$) 
of the perturbing operator $Y_5= 1 - 2 y^a y^a + \ldots$   in \rf{51} 
is then      
\be
\begin{aligned}
  \kk_{_{\rm UV}} &=0: \qquad&
 \Delta_{_{\rm UV}} &= 2-2   + \OO\Big({1\ov N}\Big)= \OO\Big({1\ov N}\Big)  \ ;
\\
\kk_{_{\rm IR} }&=- {1\ov 2}N  : \qquad&
\Delta_{_{\rm IR}}&= 2+ 2   + \OO\Big({1\ov N}\Big)= 4+ \OO\Big({1\ov N}\Big) \ .
\end{aligned}
\ee
Since $y^a$   corresponds to $\Delta=2$, the value of $\Delta_+ =4 +\ldots$ 
is consistent  with the leading-order dimension of the composite $y^a y^a $ operator. 
To  go  beyond the leading order one would need  to include higher order terms in the 
3d action \rf{39}.

\section{Defect 4-point correlation  functions  at large $N$ from M2-brane action} 
\label{sec:4pt}


Here we  follow  the same strategy as in the  case of the   correlators on the BPS Wilson line in \ci{Giombi:2017cqn} 
and  compute the  tree-level (large $N$) 4-point functions of the bosonic fields $X^I= (\xx^i, y^a)$ 
(representing the displacement operator  and the 4 scalars other than the one coupled to the surface operator in the 6d theory)
directly from the   M2-brane action \rf{35}--\rf{39}. 

Let us first   discuss  the normalization of the two-point functions. 
Given a scalar action  
\be  S_0=\frac{  \rT_p}{2} \int  d^{p+1}  x \, {\sqrt{g}}(\partial^\mu  X \partial_\mu  X +m^2  X^2) \ , \la{000}\ee
in AdS$_{p+1}$  with the metric $ds^2 = {1\ov \xxx^2} ( d\xxx^2 + d \vx d \vx ) $  (cf. \rf{34})
the  bulk-to-boundary propagator  will be  normalized as in \cite{Freedman:1998tz} (here $x=(z,\vx)$), i.e.\foot{Note that in 
\ci{Giombi:2017cqn,Beccaria:2019dws} a different normalization was used:  $C_\Delta \to   {\cal C}_\Delta= 
\frac{\Gamma(\Delta)}{2\pi^{\frac{p}{2}}\Gamma(\Delta-\frac{p}{2} +1 )}$.} 
\begin{equation}\la{111}
G^{\Delta}_{B\partial}(x,\vx')= C_\Delta \Big[\frac{\xxx }{\xxx ^2+(\vec{x}-\vec{x}')^2}\Big]^\Delta\;,\qquad 
C_\Delta= \frac{\Gamma(\Delta)}{\pi^{{p}\ov {2}}\Gamma(\Delta-\frac{p}{2})} \ ,  \qquad m^2=\Delta(\Delta-p)\ . 
\end{equation}
and the   two-point function of the corresponding boundary operator  $\OO(\vx)$   will be  defined as
\begin{equation}\la{222}
\llangle \mathcal{O}(\vx_1)\, \mathcal{O}(\vx_2)\rrangle\equiv \langle X(\vx_1)\, X(\vx_2)\rangle=
\frac{\rC_X}{|\vx_{12}|^{2\Delta}}\;, \qquad 
\rC_X= \rT_p\, (2\Delta-p) \, C_\Delta\, .
\end{equation}
In the case of the scalars $\xx^i$ and $y^a$ in \rf{36}  with masses $m^2_\xx=p+1$  and $m^2=0  $  
corresponding to $\Delta_\xx= p+1$ and $\Delta_y= p$ 
respectively  we thus find 
\be
\begin{aligned} \la{333}
\langle \xx^i (\vx_1)\, \xx^j (\vx_2)\rangle&= \delta^{ij}
\frac{\rC_\xx }{|\vx_{12}|^{2p+2 }}\ , \qquad & 
\rC_\xx &= {2 \Big(1+{2\ov p}\Big)} \,  \rC_y \ , 
\\
\langle y^a (\vx_1)\,  y^b  (\vx_2)\rangle&= \delta^{ab} 
\frac{\rC_y }{|\vx_{12}|^{2p }}\ , 
&   
\rC_y&=  \rT_p  \frac{ \Gamma(p+1)}{ \pi^{p\ov 2} \Gamma ( {p\ov 2}) }\ . 
\end{aligned} 
\ee
In particular, in the case of the string in AdS$_5\times S^5$    ($\rT_1= {\sqrt\lambda \ov 2 \pi} $) we find
$ \rC_\xx = 6 \rC_y, \ \rC_y= {\sqrt\lambda \ov 2 \pi^2}$
(in agreement with direct identification  of  $\xx^i$   with displacement operator 
and $y^a$  with ``transverse" scalars   and strong-coupling limit of exact prediction in
\cite{Drukker:2011za,Correa:2012at}).
In the present case of $p=2$, $\TT= { 2 N \ov \pi}$  we  find\foot{Note that  here we start  with \rf{36}, i.e.
assume that the scalars $y^a$ are normalized in 
the same way  as $\xx^i$  (after the rescaling  $y^a$  by $\rr=\ha$, cf. \rf{45}).
If the corresponding   scalar  operator  on the defect  is  identified with unrescaled $y^a$  we get 
$ \rC_y\to \rr^2 \rC_y =  {  N\ov 2 \pi^2}$  while  $\rC_\xx$ is  the same. 
See also \cite{Drukker:2020atp}.
}
\be \la{555}
\rC_\xx = { 16 N\ov \pi^2}  =  4 \rC_y \ , \qquad \qquad   \rC_y=  { 4 N\ov \pi^2}   \ . 
\ee
Since the $\xx^i$ fluctuations are dual to the displacement operator $\DD^i$, from the above results we can read off the normalization $\rC_{\DD}$ of the $\DD^i$ two-point function on the surface defect in the (2,0) theory
\be \la{666}
\llangle \DD^i(\vx_1) \DD^j(\vx_2)\rrangle = \delta^{ij} \frac{\rC_{\DD}}{\vx_{12}^6}\,,\qquad \rC_{\DD} = \frac{16N}{\pi^2}+{\cal O}(N^0)\,.
\ee  
The normalization constant $\rC_\DD$ also determines the anomaly coefficient associated to 
extrinsic curvature of the surface \cite{bianchi:2015liz}.

\subsection{The $\langle yyyy \rangle$ correlator}\label{sec:yyyy}

Let us first  compute  the 4-point  boundary correlator of the four $S^4$ fluctuations
\begin{equation}
\label{777}
G(\vx_i,\tt_i)=\langle y(\vx_1;\tt_1)\,y(\vx_2;\tt_2)\,y(\vx_3;\tt_3)\, y(\vx_4;\tt_4)\rangle \ , 
\end{equation}
where we have multiplied each $y^a(\vx_i)$ with an auxiliary constant 4-vector $\tt^a_i$ to remove the
$SO(4)$   R-symmetry indices
\begin{equation}
y(\vx;\tt)\equiv   \tt^a\, y^a(\vx)\;.\la{420}
\end{equation}
Here $\vx$ stands for  2 boundary coordinates of AdS$_3$.
It is convenient to extract a kinematic factor so that the correlation function can be expressed in terms of the cross ratios
\begin{equation}\la{430}
G=\frac{\tt_{12}\tt_{34}}{\vx_{12}^4\vx_{34}^4}\, \mathcal{G}(\ze,\bar{\ze};\alpha,\bar{\alpha})\ , 
\end{equation}
where $\tt_{ij}\equiv \tt_i\cdot \tt_j$, and 
\begin{equation}
U=\frac{\vx_{12}^2\vx_{34}^2}{\vx_{13}^2\vx_{24}^2}=\ze\bar{\ze}\;,\quad V=\frac{\vx_{14}^2\vx_{23}^2}{\vx_{13}^2\vx_{24}^2}=(1-\ze)(1-\bar{\ze})\;,\quad  \sigma=\frac{\tt_{13}\tt_{24}}{\tt_{12}\tt_{34}}=\alpha\bar{\alpha}\;,\quad \tau=\frac{\tt_{14}\tt_{23}}{\tt_{12}\tt_{34}}=(1-\alpha)(1-\bar{\alpha})\;.\la{44}
\end{equation}
The holographic computation of the correlator leads to a $1/N$ expansion 
\begin{equation}\la{4.5}
\mathcal{G}=\mathcal{G}_{\rm disc}+\mathcal{G}_{\rm tree}+\ldots\;, 
\end{equation}
where the leading  order 
contribution   is given by the  disconnected diagrams
\begin{equation}\label{Gdisc}
\mathcal{G}_{\rm disc}=\frac{16N^2}{\pi^4}\, \Big(1+\sigma U^2+\tau \frac{U^2}{V^2}\Big)\;.
\end{equation}
Note that this scales as   $(\rC_y)^2$ in  agreement with \rf{333},\rf{555}.
The order $N$  tree-level contributions to the 4-point function can be divided into two parts: $\mathcal{G}_1$ from the Dirac-Nambu type action \rf{35},\rf{39}   which is parity-even, 
and $\mathcal{G}_2$ from WZ type action
\rf{399},\rf{40}  which is parity odd
\begin{equation}\la{4.7}
\mathcal{G}_{\rm tree}=\mathcal{G}_1+\mathcal{G}_2\;.
\end{equation}
The first contribution can be straightforwardly computed from the action, and after some simplification reads
(after using \rf{2.7})
\begin{equation}\la{488}
\begin{split}
\mathcal{G}_1&=-\frac{96N}{\pi^4}\,  U^2\bigg(\Big[ (U-1-V)\bar{D}_{3333}-U\bar{D}_{3322}+\bar{D}_{2222}\Big]
+\sigma \Big[(1-U-V)\bar{D}_{3333}-\bar{D}_{3232}+\bar{D}_{2222}\Big]\\
{}&\qquad\qquad  +\tau\Big[(V-1-U)\bar{D}_{3333}-\bar{D}_{3223}+\bar{D}_{2222}\Big] \bigg)\;.
\end{split}
\end{equation}
Here $\bar D_{\Delta_1\Delta_2\Delta_3\Delta_4}$ are functions of  cross-ratios, and are related to the $D$-functions defined by
\begin{equation}\la{4160}
D_{\Delta_1\Delta_2\Delta_3\Delta_4}(\vec{x}_i)=\int \frac{d\xxx \,d^{p}\vec{x}}{\xxx ^{p+1}}\prod_{i=1}^4 \Big(\frac{\xxx }{\xxx ^2+(\vec{x}-\vec{x}_i)^2}\Big)^{\Delta_i}\;,
\end{equation}
via 
\begin{equation}\la{4170}
\frac{2}{\pi^{\frac{d}{2}}} \frac{\prod_{i=1}^4\Gamma(\Delta_i)}{\Gamma(\Sigma-\frac{1}{2}d)}\, D_{\Delta_1\Delta_2\Delta_3\Delta_4}=\frac{(\vec{x}_{14}^2)^{\Sigma-\Delta_1-\Delta_4}(\vec{x}_{34}^2)^{\Sigma-\Delta_3-\Delta_4}}{(\vec{x}_{13}^2)^{\Sigma-\Delta_4}(\vec{x}_{24}^2)^{\Delta_2}}\bar{D}_{\Delta_1\Delta_2\Delta_3\Delta_4}\;, \quad\quad \Sigma=\half \sum_{i=1}^4\Delta_i\;.
\end{equation}

The three-derivative contact Witten diagram  corresponding to the WZ  vertex 
\rf{40} has already been computed in \cite{Rastelli:2019gtj}, and with the normalization of the WZ term in (\ref{40}) we have 
\begin{equation}\la{49}
\mathcal{G}_2=-\frac{72N}{\pi^4}\,  U^2(\ze-\bar{\ze})(\alpha-\bar{\alpha})\bar{D}_{3333}\;.
\end{equation}

Note that  the combination \rf{4.7} 
of  the two contributions \rf{488}  and \rf{49}  is very special, in that the resulting  correlator satisfies differential relations which resemble superconformal Ward identities (see, {\it e.g.}, \cite{Dolan:2004mu} for examples in SCFT$_d$ with $d>2$)
\begin{equation}\label{scfwardid}
\Big({-\frac{1}{2}}\ze\partial_\ze+\alpha\partial_\alpha\Big)\mathcal{G}(\ze,\bar{\ze};\alpha,\bar{\alpha})\Big|_{\alpha=1/\ze}=0\;,\qquad \qquad 
\Big({-\frac{1}{2}}\bar{\ze}\partial_{\bar{\ze}}+\bar{\alpha}\partial_{\bar{\alpha}}\Big)\mathcal{G}(\ze,\bar{\ze};\alpha,\bar{\alpha})\Big|_{\bar{\alpha}=1/\bar{\ze}}=0\;.
\end{equation}
Here it is understood that we first act with the differential operators on the correlator, and then set the R-symmetry cross ratios to specific values. One can easily check that the disconnected correlator $\mathcal{G}_{\rm disc}$ in \rf{Gdisc} 
also satisfies the identities above. We will show in Section~\ref{sec:super} below that (\ref{scfwardid}) are indeed Ward identities following from the superconformal symmetry $[OSp(4^*|2)]^2$. 

We note that the same differential identities are  satisfied by the 
correlators on the  $1\ov  2$-BPS Wilson loop in the string theory case (see  Appendix~\ref{sec:WL}). The major difference,  however, 
is that in the 1d Wilson loop case there is only a single conformal cross ratio $\ze$, 
and both $\alpha$ and $\bar{\alpha}$ can be twisted with respect to $\ze$. 
Here the parity odd term \rf{49}
has broken the  symmetry of interchanging $\ze$ and $\bar{\ze}$ in the correlator: only the simultaneous interchange of $\ze\leftrightarrow \bar{\ze}$, $\alpha\leftrightarrow \bar{\alpha}$ remains a symmetry of the 4-point function. Therefore,  we have two differential identities which separately pair up $\ze$ with $\alpha$ and $\bar{\ze}$ with $\bar{\alpha}$, and the chirality of these relations parallels the factorized structure of the superconformal group $OSp(4^*|2)\times OSp(4^*|2)$.  The same structure of 
the superconformal Ward identities was observed in \cite{Rastelli:2019gtj}  for $PSU(1,1|2)\times PSU(1,1|2)$.

\subsection{The $\langle {\rm x} {\rm x} {\rm x} {\rm x} \rangle$ and $\langle {\rm x} {\rm x} yy \rangle$ correlators and Mellin representation}

The calculation for the $\langle {\rm x} {\rm x} {\rm x} {\rm x} \rangle$ and $\langle {\rm x} {\rm x} yy \rangle$ correlators is almost identical to that of the  the parity-even part of the $\langle yyyy\rangle$ correlator  as 
they are not affected by the WZ term at tree level, and only the Dirac-Nambu term  \rf{35} contributes.
We  first  present 
the results for generic defect dimension $p$  and then specify to $p=2$.  For  the string (Wilson loop) case of $p=1$ 
the expressions  below agree with the results of \ci{Giombi:2017cqn}. 

Using  the action in \rf{35}-\rf{39}
we find for  the 4-point function of  the transverse AdS$_{d+1}$ fluctuations 
\begin{equation}
\begin{split}
{}&\langle  {\rm x}^i(\vec{x}_1) {\rm x}^j(\vec{x}_2) {\rm x}^k(\vec{x}_3) {\rm x}^l(\vec{x}_4)\rangle=- \frac{{\rm T}_p\, (1+p)^2\Gamma^4(1+p)}{\pi^{2p}\Gamma^4(1+\frac{p}{2})}\bigg(\delta^{ij}\delta^{kl}\Big[ (2+p)(4+5p)D_{p+1,p+1,p+1,p+1}\\
{}&-4(2+p)(1+2p)\vec{x}_{34}^2D_{p+1,p+1,p+2,p+2}-4(1+p)^2(\vec{x}_{14}^2\vec{x}_{23}^2+\vec{x}_{13}^2\vec{x}_{24}^2-\vec{x}_{12}^2\vec{x}_{34}^2)D_{p+2,p+2,p+2,p+2}\Big] \\
{}&+\delta^{ik}\delta^{jl}\Big[ (2+p)(4+5p)D_{p+1,p+1,p+1,p+1}-4(2+p)(1+2p)\vec{x}_{24}^2D_{p+1,p+2,p+1,p+2}\\
{}&-4(1+p)^2(\vec{x}_{14}^2\vec{x}_{23}^2+\vec{x}_{12}^2\vec{x}_{34}^2-\vec{x}_{13}^2\vec{x}_{24}^2)D_{p+2,p+2,p+2,p+2}\Big] \\
{}&+\delta^{il}\delta^{jk}\Big[(2+p)(4+5p)D_{p+1,p+1,p+1,p+1}-4(2+p)(1+2p)\vec{x}_{23}^2D_{p+1,p+2,p+2,p+1}\\
{}&-4(1+p)^2(\vec{x}_{12}^2\vec{x}_{34}^2+\vec{x}_{13}^2\vec{x}_{24}^2-\vec{x}_{14}^2\vec{x}_{23}^2)D_{p+2,p+2,p+2,p+2}\Big] \bigg)\ , 
\end{split}
\end{equation}
where we have already used the relation $\Delta_{\rm x}=p+1$, and the $D$-function identities (summarized in, {\it e.g.}, Appendix~D of \cite{Arutyunov:2002fh}) to simplify the expression. 
For our present  case, $p=2, \ \TT= {2N\ov \pi}$,  this  correlator  may be  written explicitly 
in terms of the  $\bar D$-functions of cross ratios as (cf. \rf{488})
\begin{equation}
\begin{split}
\langle  {\rm x}^i(\vec{x}_1) {\rm x}^j(\vec{x}_2) {\rm x}^k(\vec{x}_3) {\rm x}^l(\vec{x}_4)\rangle&=-\frac{182 N}{\pi^4 \vec{x}_{12}^6\vec{x}_{34}^6}U^3\bigg(\delta^{ij}\delta^{kl}\Big[63 \bar{D}_{3333}-50\bar{D}_{3344}+15(U-V-1)\bar{D}_{4444}\Big]\\
{}&\qquad \qquad \qquad +\delta^{ik}\delta^{jl}\Big[63 \bar{D}_{3333}-50\bar{D}_{3434}+15(1-U-V)\bar{D}_{4444}\Big]\\
{}&\qquad \qquad \qquad +\delta^{il}\delta^{jk}\Big[63 \bar{D}_{3333}-50\bar{D}_{3443}+15(V-1-U)\bar{D}_{4444}\Big]\bigg)\;.
\end{split}
\end{equation}
Similarly, the $\langle {\rm x} {\rm x} yy \rangle$ correlator for generic $p$ reads 
\begin{equation}
\begin{split}
\langle {\rm x}^i (\vec{x}_1){\rm x}^j (\vec{x}_2)& y^a (\vec{x}_3) y^b (\vec{x}_4)\rangle=-\frac{{\rm T}_p\, 4^{2p-1}\Gamma^4(\frac{1+p}{2})}{p^2\pi^{2p+2}}\delta^{ij}\delta^{ab}\bigg({-(2+p+p^2)}D_{1+p,1+p,p,p}\\{}&-2p(3+p)\vec{x}_{34}^2D_{p+1,p+1,p+1,p+1}+2(1+p)^2\Big[ {-\vec{x}_{12}^2} D_{p+2,p+2,p,p}+\vec{x}_{13}^2 D_{p+2,p+1,p+1,p}\\
{}&+\vec{x}_{14}^2 D_{p+2,p+1,p,p+1}+\vec{x}_{23}^2 D_{p+1,p+2,p+1,p}+\vec{x}_{24}^2 D_{p+1,p+2,p,p+1}\\
{}&-2(\vec{x}_{14}^2\vec{x}_{23}^2+\vec{x}_{13}^2\vec{x}_{24}^2-\vec{x}_{12}^2\vec{x}_{34}^2)D_{p+2,p+2,p+1,p+1}\Big] \bigg)\ , 
\end{split}
\end{equation}
where we have used that $\Delta_{\rm x}=p+1$ and $\Delta_{y}=p$. For the $p=2$
the correlator may be written in terms of  functions of cross ratios as 
(after using again  the $D$-function identities)
\begin{equation}
\begin{split}
{}&\langle {\rm x}^i (\vec{x}_1){\rm x}^j (\vec{x}_2) y^a (\vec{x}_3) y^b (\vec{x}_4)\rangle=-\frac{96N}{\pi^4}\frac{\delta^{ij}\delta^{ab}}{\vec{x}_{12}^6\vec{x}_{34}^4}U^3\Big[5(U-1-V)\bar{D}_{4433}-13\bar{D}_{3333}+8\bar{D}_{3322}\Big]\;.
\end{split}
\end{equation}
For comparison, let us also record
the parity-even part of the $\langle yyyy \rangle$ correlator with general $p$ and $\rr$ 
following from \rf{39} (generalizing the $p=2, \ \rr=\ha$ expression in \rf{430},\rf{4.7},\rf{488}) 
\begin{equation}
G_1(\vx_i,\tt_i)=-\frac{{\rm T}_p\ p^2 \Gamma^4(p)}{\pi^{2p}\Gamma^4(\frac{p}{2})}\Big[\tt_{12}\tt_{34} G_1^{12;34}(\ze,\bar{\ze})+\tt_{13}\tt_{24} G_1^{13;24}(\ze,\bar{\ze}))+\tt_{14}\tt_{23} G_1^{14;23}(\ze,\bar{\ze})\Big]\ , \la{415}
\end{equation}
\begin{equation}
\begin{split}
\no G_1^{12;34}&=-\tfrac{2}{\rr^2}(D_{p,p,p,p}-2\vec{x}_{12}^2D_{p+1,p+1,p,p})+p^2(5D_{p,p,p,p}-8\vec{x}_{12}^2D_{p+1,p+1,p,p})\\
{}&\quad{}-4p^2(\vec{x}_{14}^2\vec{x}_{23}^2+\vec{x}_{13}^2\vec{x}_{24}^2-\vec{x}_{12}^2\vec{x}_{34}^2)D_{p+1,p+1,p+1,p+1}\;,
\end{split}
\end{equation}
\begin{equation}
\begin{split}
\no G_1^{13;24}&=-\tfrac{2}{\rr^2}(D_{p,p,p,p}-2\vec{x}_{13}^2D_{p+1,p,p+1,p})+p^2(5D_{p,p,p,p}-8\vec{x}_{13}^2D_{p+1,p,p+1,p})\\
{}&\quad{}-4p^2(\vec{x}_{14}^2\vec{x}_{23}^2+\vec{x}_{12}^2\vec{x}_{34}^2-\vec{x}_{13}^2\vec{x}_{24}^2)D_{p+1,p+1,p+1,p+1}\;,
\end{split}
\end{equation}
\begin{equation}
\begin{split}
\no G_1^{14;23}&=-\tfrac{2}{\rr^2}(D_{p,p,p,p}-2\vec{x}_{14}^2D_{p+1,p,p,p+1})+p^2(5D_{p,p,p,p}-8\vec{x}_{14}^2D_{p+1,p,p,p+1})\\
{}&\quad{}-4p^2(\vec{x}_{12}^2\vec{x}_{34}^2+\vec{x}_{13}^2\vec{x}_{24}^2-\vec{x}_{14}^2\vec{x}_{23}^2)D_{p+1,p+1,p+1,p+1}\;,
\end{split}
\end{equation}
$G_1$ in \rf{415} is related to $\mathcal{G}_1$ defined in (\ref{4.7})  as in \rf{430}, i.e.  
\begin{equation}
G_1(x_i,\tt_i)=\frac{\tt_{12}\tt_{34}}{\vec{x}_{12}^{2p}\vec{x}_{34}^{2p}}\mathcal{G}_1(\ze,\bar{\ze};\alpha,\bar{\alpha})\;.
\end{equation}
Note that  the WZ term in \rf{399},\rf{40} is specific to $p=2$ case so that its contribution to the 4-point correlator \rf{49} does not admit  a generalization to an arbitrary $p$.

We can develop a better intuition about the  above  expressions for the correlators 
using the Mellin representation  \cite{Mack:2009mi,Penedones:2010ue}.  It is straightforward to translate the 
$D$-functions into the Mellin space (see, {\it e.g.}, Appendix~A of  \cite{Zhou:2018sfz} for 
explicit expressions), and we find ($s+t+u=\sum_r \Delta_r$)
\begin{align}
\la{417}
&\langle  {\rm x}^i(\vec{x}_1) {\rm x}^j(\vec{x}_2) {\rm x}^k(\vec{x}_3) {\rm x}^l(\vec{x}_4)\rangle
=\frac{F_1(U,V)}{\vec{x}_{12}^{2p+2}\vec{x}_{34}^{2p+2}}\ , \\
& F_1(U,V)= 
\int_{-i\infty}^{i\infty}\frac{dsdt}{(4\pi i)^2}U^{\frac{s}{2}}V^{\frac{t-2p-2}{2}}\ 
\mathcal{M}
^{ijkl}(s,t)\ \Gamma^2(\tfrac{2p+2-s}{2})\Gamma^2(\tfrac{2p+2-t}{2})\Gamma^2(\tfrac{2p+2-u}{2})\;,
\quad s+t+u= 4p + 4 \ ,  \no  \\
&\langle {\rm x}^i (\vec{x}_1){\rm x}^j (\vec{x}_2) y^a (\vec{x}_3) y^b (\vec{x}_4)\rangle=\frac{F_2(U,V)}{\vec{x}_{12}^{2p}\vec{x}_{34}^{2p+2}}\ ,\la{419}  \\
& F_2(U,V) = \int_{-i\infty}^{i\infty}\frac{dsdt}{(4\pi i)^2}U^{\frac{s}{2}}V^{\frac{t-2p-1}{2}}\ \mathcal{M} 
^{ij;ab}(s,t)\ \Gamma(\tfrac{2p-s}{2})\Gamma(\tfrac{2p+2-s}{2})\Gamma^2(\tfrac{2p+1-t}{2})\Gamma^2(\tfrac{2p+1-u}{2}) \ ,\ \   s+t+u=4p +2 \no 
\end{align}
where the Mellin amplitudes are given by 
\begin{align}
&\mathcal{M} 
^{ijkl}(s,t)={}\frac{{\rm T}_p\,  \Gamma \big(\frac{3 p}{2}+4\big)}{3\pi ^{\frac{3 p}{2}}(3 p+4) \Gamma^4 (\frac{p}{2}+1)}
\Big[\delta^{ij}\delta^{kl} M(t,u) +  \delta^{ik}\delta^{jl} M(s,u) + \delta^{il}\delta^{jk} M(s,t) \Big]   \;,
\\
&M(t,u) = -3 (3 p+4)tu +2 (2 p+1) (3 p+4)(t+u)-4 p (p+1) (4 p+5)\ \stackrel{p=2}{\to} \   -30 tu + 100 (t + u) -312   \ , \no\\
\la{yyxx}
&\mathcal{M}
^{ij;ab}(s,t)={}\frac{  {\rm T}_p\ p^2   \Gamma \big(\frac{3 p}{2}+1\big)}{16 \pi ^{\frac{3 p}{2}} \Gamma^4 \big(\frac{p}{2}+1\big)}
\ \delta^{ij}\delta^{ab}   M' (t,u) \ ,  \\
&  M'(t,u) =-(3 p+2) (3 p+4)tu+p (3 p+2) (4 p+5)(t+u)
-16 p^4-28 p^3-5 p^2+14 p+8 \no \\ 
&\qquad \quad    \  \stackrel{p=2}{\to}  \   - 80 tu + 208 (t + u) -464 \ .   \no   
\end{align}
Similarly, the parity-even part of the $\langle yyyy \rangle$  correlator \rf{415} admits the 
following Mellin representation
\begin{align}\no
&G_1(\vx_i,\tt_i)=\frac{1}{\vec{x}_{12}^{2p}\vec{x}_{34}^{2p}}\int_{-i\infty}^{i\infty}\frac{dsdt}{(4\pi i)^2}U^{\frac{s}{2}}V^{\frac{t-2p}{2}}\ 
\mathcal{M}_1(s,t;\tt_i)\ \Gamma^2(\tfrac{2p+2-s}{2})\Gamma^2(\tfrac{2p-t}{2})\Gamma^2(\tfrac{2p-u}{2})\ , 
\quad  s+t+u=4p \ , \\
\la{yyyyeven}
&\mathcal{M}_1(s,t;\tt_i)=-\frac{{\rm T}_p\, \Gamma(\frac{3p}{2}+1)}{6\pi^{\frac{3p}{2}}\, \Gamma^4(\frac{p}{2})}\Big[\tt_{12}\tt_{34}M''(t,u)+\tt_{13}\tt_{24}M''(s,u)+\tt_{14}\tt_{23}M''(s,t)\Big]\;,\\
&\no M''(t,u)=3(2+3p) tu-6(2p^2-\tfrac{1}{\rr^2} )(t+u)+4p(4p^2-3p-\tfrac{4}{\rr^2})\quad {\stackrel{p=2,\, \rr={1\ov 2} }{\to}}\quad 6tu-6(t+u)-12\;.
\end{align}
As expected, the contact interactions with up to four derivatives (cf.  \rf{36}--\rf{39})
give amplitudes that are quadratic polynomials in the Mellin-Mandelstam variables. 

Let us also point out that the parity-odd part  \rf{49} 
of the $\langle yyyy \rangle$ correlator, which only exists for 
$p=2$ as it  derives from the WZ action (\ref{399}), 
does  not  admit a Mellin representation.  Indeed, 
the parity-odd contribution (\ref{49}) 
gains a minus sign under $\ze\leftrightarrow \bar{\ze}$ 
while the variables $U$ and $V$  (that appear in the standard  definition of the 
Mellin representation) are invariant under this  transformation (cf. their definition in \rf{44}).

\subsection{OPE analysis}
\label{sec:OPE}

In this subsection, we perform a preliminary OPE analysis to extract the CFT data for low-lying operators from the 4-point correlator $\langle yyyy \rangle$. Because of the chiral nature of the correlator, we have to use 2d conformal blocks which are not symmetrized with respect to $\ze$ and $\bar{\ze}$
\begin{equation}
g_{h,\bar{h}}(\ze,\bar{\ze})=\ze^{\frac{h}{2}}\bar{\ze}^{\frac{\bar{h}}{2}} \,  {}_2F_1(\tfrac{h}{2},\tfrac{h}{2},h;\ze)\ {}_2F_1(\tfrac{\bar{h}}{2},\tfrac{\bar{h}}{2},\bar{h};\bar{\ze})\;,\qquad \qquad h-\bar{h}\in \mathbb{Z}\ . 
\end{equation}
This conformal block corresponds to an operator with holomorphic dimension $h$ and anti-holomorphic dimension $\bar{h}$.\footnote{They are related to $\Delta$ and $\ell$ by $\min\{h,\bar{h}\}=\Delta-\ell$,\  $\max\{h,\bar{h}\}=\Delta+\ell$.} For the same reason, we should use $SU(2)$ R-symmetry polynomials for $SU(2)_L\times SU(2)_R=SO(4)$
\begin{equation}
R_{L,m}(\alpha)=P_m(1-2\alpha)\;,\qquad R_{R,m}(\bar{\alpha})=P_m(1-2\bar{\alpha})\ , 
\end{equation}
where $m=0,1,\cdots$ corresponds to the spin-$\frac{m}{2}$ representation of $SU(2)$.

We can analyze the 4-point function in small $\ze$ (or $\bar{\ze}$) expansion,\footnote{The analysis of small $\ze$ expansion is identical to  that of the small $\bar{\ze}$ expansion, after interchanging $\alpha$ and $\bar{\alpha}$. Therefore,  in the following we will only focus on the former.}
which is  dominated by operators with small $h$ (or $\bar{h}$). By going to higher orders in the expansion, we can systematically read off CFT data for operators with increasing conformal twists. 
However, at higher conformal twists the mixing effect of 
operators  become  important, and analyzing the $\langle yyyy\rangle$ correlator 
alone  gives only  ``averages'' of
anomalous dimensions over the operators appearing in the mixing (weighted by the corresponding
OPE coefficients).

We will postpone the unmixing analysis (which involves $\langle {\rm x} {\rm x} {\rm x} {\rm x} \rangle$ and $\langle {\rm x} {\rm x} y y \rangle$ as well) for a future study and 
focus on the leading double-trace operators with $h=4$ where the mixing is absent. 
Since we are only interested in  the leading operators in the OPE, we do not need superconformal blocks and bosonic conformal blocks are sufficient. 

It is useful to decompose the correlator \rf{4.5}  into different R-symmetry channels
\begin{equation}\la{4.37}
\begin{split}
\mathcal{G}(\ze,\bar{\ze};\alpha,\bar{\alpha})&=R_{L,0}(\alpha)R_{R,0}(\bar{\alpha})\, \mathcal{G}_{(\mathbf{1},\mathbf{1})}(\ze,\bar{\ze})+R_{L,1}(\alpha)R_{R,0}(\bar{\alpha})\, \mathcal{G}_{(\mathbf{2},\mathbf{1})}(\ze,\bar{\ze})\\
{}&\quad{}+R_{L,0}(\alpha)R_{R,1}(\bar{\alpha})\, \mathcal{G}_{(\mathbf{1},\mathbf{2})}(\ze,\bar{\ze})+R_{L,1}(\alpha)R_{R,1}(\bar{\alpha})\, \mathcal{G}_{(\mathbf{2},\mathbf{2})}(\ze,\bar{\ze})\ , 
\end{split}
\end{equation}
where $\mathbf{R}= (\mathbf{1},\mathbf{1}), \ldots$ in $\mathcal{G}_{\mathbf{R}}(\ze,\bar{\ze})$ labels the  representation of $SU(2)_L\times SU(2)_R$ that is exchanged. 
Each $\mathcal{G}_{\mathbf{R}}(\ze,\bar{\ze})$ can be decomposed into the conformal blocks
\begin{equation}\la{4.38}
\mathcal{G}_{\mathbf{R}}(\ze,\bar{\ze})=\underbrace{\sum_{h,\bar{h}} C^{(0),\mathbf{R}}_{h,\bar{h}} g_{h,\bar{h}}(\ze,\bar{\ze})}_{\rm disconnected}+\underbrace{\sum_{h,\bar{h}} C^{(1),\mathbf{R}}_{h,\bar{h}} g_{h,\bar{h}}(\ze,\bar{\ze})+\frac{1}{2}\gamma^{(1),\mathbf{R}}_{h,\bar{h}} C^{(0),\mathbf{R}}_{h,\bar{h}} (\partial_h+\partial_{\bar{h}})g_{h,\bar{h}}(\ze,\bar{\ze})}_{\text{tree level}} +\ldots\ , 
\end{equation}
where $\gamma^{(1),\mathbf{R}}_{h,\bar{h}}$ are anomalous dimensions associated with $\log(z\bar{z})$ divergences in the correlator, {\it i.e.},  
\begin{equation}\la{4.39}
h_{\rm exact}=h+\gamma^{(1),\mathbf{R}}_{h,\bar{h}}+\ldots\;,\quad \bar{h}_{\rm exact}=\bar{h}+\gamma^{(1),\mathbf{R}}_{h,\bar{h}}+\ldots\;, 
\end{equation}

{\bf The $(\mathbf{1},\mathbf{1})$ channel:}
\noindent
\ \ From the $\ze^2$ coefficient in the small $\ze$ expansion of the disconnected part of the correlator  in \rf{Gdisc} 
comparing to \rf{4.38}    we  find that 
\begin{equation}
C^{(0),(\mathbf{1},\mathbf{1})}_{4,\bar{h}}=\frac{2^{4-\bar{h}} (\bar{h}-2)  \Gamma (\frac{\bar{h}}{2}+1)}{\pi ^{7/2} \Gamma (\frac{\bar{h}-1}{2})}\, N^2  \;, \qquad\qquad  \bar{h}\in 4 \mathbb{Z}_+\;.
\end{equation}
From the $\ze^2\log (\ze\bar{\ze})$ coefficient of the tree-level correlator \rf{4.7}  we can read off $\gamma^{(1),(\mathbf{1},\mathbf{1})}_{4,\bar{h}} C^{(0),(\mathbf{1},\mathbf{1})}_{4,\bar{h}}$.  We find   contributions  only 
from $\bar{h}=4,8$, which correspond to spin 0 and spin 2 operators respectively. As a result, 
\begin{equation}
\gamma^{(1),(\mathbf{1},\mathbf{1})}_{4,4}=-\frac{24}{5N}\;,\qquad\qquad  \gamma^{(1),(\mathbf{1},\mathbf{1})}_{4,8}=-\frac{48}{35N}\;.
\end{equation}
The fact that the anomalous dimensions have a finite support on spins is expected because the tree-level correlator is computed as a finite sum of contact diagrams, and each contact diagram has a finite support on spins \cite{Heemskerk:2009pn}.

\paragraph{ The $(\mathbf{1},\mathbf{2})$ channel:}
From the $\ze^2$ expansion coefficient in  the disconnected part   \rf{Gdisc}  we find 
\begin{equation}
C^{(0),(\mathbf{1},\mathbf{2})}_{4,\bar{h}}=\frac{2^{4-\bar{h}} (\bar{h}-2)  \Gamma (\frac{\bar{h}}{2}+1)}{\pi ^{7/2} \Gamma (\frac{\bar{h}-1}{2})}\, N^2  \;, \qquad\qquad  \bar{h}\in 4 \mathbb{Z}_++2\;.
\end{equation}
From  the $\ze^2\log (\ze\bar{\ze})$ coefficient in  the tree-level correlator  \rf{4.7}  
we can extract the  corresponding anomalous dimension. We find that there is only one operator with $\bar{h}=6$ 
contributing and  its  anomalous dimension is 
\begin{equation}
\gamma^{(1),(\mathbf{1},\mathbf{2})}_{4,6}=-\frac{24}{5N}\;.
\end{equation}

\paragraph{The $(\mathbf{2},\mathbf{1})$ channel:}
From the small $\ze$ expansion of the  disconnected correlator   we find again
\begin{equation}
C^{(0),(\mathbf{2},\mathbf{1})}_{4,\bar{h}}=\frac{2^{4-\bar{h}} (\bar{h}-2)  \Gamma (\frac{\bar{h}}{2}+1)}{\pi ^{7/2} \Gamma (\frac{\bar{h}-1}{2})}N^2 \;, \qquad\qquad  \bar{h}\in 4 \mathbb{Z}_++2\;.
\end{equation}
However,  here there is no $\ze^2\log(\ze\bar{\ze})$  term  in the tree-level correlator. 
The operator with $h=4$ and $\bar{h}=6$ receives only a correction to the OPE coefficient
\begin{equation}
C^{(1),(\mathbf{2},\mathbf{1})}_{4,6}=-\frac{4}{5\pi^4}\, N \;.
\end{equation}
The first logarithmic singularity arises at $\ze^3$ order which corresponds to $h=6$. There are two operators responsible for this logarithmic singularity, with the anti-holomorphic dimensions $\bar{h}=4,8$.

\paragraph{The $(\mathbf{2},\mathbf{2})$ channel:}
Here the zeroth order OPE coefficients read
\begin{equation}
C^{(0),(\mathbf{2},\mathbf{2})}_{4,\bar{h}}=\frac{2^{4-\bar{h}} (\bar{h}-2)  \Gamma (\frac{\bar{h}}{2}+1)}{\pi ^{7/2} \Gamma (\frac{\bar{h}-1}{2})}\, N^2 \;, \qquad\qquad  \bar{h}\in 4 \mathbb{Z}_+\;.
\end{equation}
As in the $(\mathbf{2},\mathbf{1})$ channel  here  we find 
no    $\ze^2\log(\ze\bar{\ze})$  term 
in the tree-level correlator. 
This is consistent with the fact that the $(\mathbf{2},\mathbf{2})$ channel receives contribution from the 
$1\ov  2$-BPS operator with $h=\bar{h}=4$  that  has no anomalous dimension. The tree-level correlator leads only to a correction to
the OPE coefficient of  this  operator 
\begin{equation}
C^{(1),(\mathbf{2},\mathbf{2})}_{4,4}=-\frac{4}{\pi^4} N \;.
\end{equation}

\section{Superconformal symmetry of holographic correlators}
\label{sec:super}
In Section~\ref{sec:yyyy} we observed that the $\langle yyyy\rangle$ correlator satisfies intricate differential relations (\ref{scfwardid}) 
which can be interpreted as superconformal Ward identities following from the global symmetry $[D(2,1|-\frac{1}{2})]^2=[OSp(4^*|2)]^2$. We give a derivation of these identities in Section~\ref{sec:twisted4pt} by studying the supersymmetry of 
``twisted'' operators defined below. 
In fact, these superconformal Ward identities are so restrictive to essentially fix the form of the 4-point function. 
This is shown in Section~\ref{sec:ward}, where we ``bootstrap'' the $\langle yyyy\rangle$ tree-level correlator 
from the Ward identities and crossing symmetry without inputting the precise form of the AdS$_3$ ``bulk" 
vertices in the M2 brane action, thus shortcutting the computation in Section~\ref{sec:yyyy}.

\subsection{Twisted operators and supersymmetry}\label{sec:twisted}

It is possible to construct a set of ``twisted'' combinations of the scalar fluctuations $y^a$, 
following a similar construction in \cite{Drukker:2009sf}.\footnote{This should not be confused with the chiral algebra twist of \cite{Beem:2014kka}. The construction of \cite{Beem:2014kka} requires twisting a subalgebra $\mathfrak{psu}(1,1|2)$ which does not fit into $D(2,1|-\frac{1}{2})$. Moreover, it will be clear that our twisted correlator has both $\ze$  and $\bar{\ze}$ dependence, rather than being chiral.} We start by rewriting 
\eqref{420} in spinor notations ($\alpha=1,2$) 
\begin{equation}
y(\vec x; \tt)=y(\zz,\bar\zz;\aa,\bar\aa)=
y_{\alpha\dot\alpha}\, a^\alpha\bar a^{\dot\alpha}\,,
\qquad
\tt^a= 
   \rho^a_{\alpha\dot\alpha}a^\alpha\bar a^{\dot\alpha}\,,
\qquad
a^\alpha=(1,\aa)\,,
\quad
\bar a^{\dot \alpha}=(1,\bar \aa)\,.
\end{equation}
We switched here to complex coordinates $\zz=x^1+ix^2$ and $\rho^a$ are 
$SO(4)$ gamma matrices ($\rho=(i\vec\sigma,I)$).
Explicitly
\begin{equation}
\tt^a(\aa, \bar\aa)= 
\{i(\aa+\bar\aa), \,\aa-\bar \aa,\,i(1-|\aa|^2),\,1+|\aa|^2\}\,.
\end{equation}
The inner product is now simply
\begin{equation}\la{442}
\tt^a(\aa_1, \bar \aa_1)\, \tt^a(\aa_2,\bar \aa_2)
= 2 |\aa_1-\aa_2|^2\,.
\end{equation}

We define the half-twisted operator by setting $\aa=\zz$
\begin{equation}\la{4522}
{\cal Y}(\zz,\bar \zz;\bar{\aa})=
  y(\zz,\bar\zz;\zz,\bar\aa)\,.
\end{equation}
We can also define a half-twist in the opposite chirallity by setting $\bar{\aa}=\bar{\zz}$, or twist both.

As $y^a$ is a massless ($\Delta=2$) field   whose 2-point function is given by \rf{333}, 
eq.\rf{442} implies that $\cal Y$ behaves like  
\begin{equation}
\left\langle{\cal Y}(\zz_1,\bar \zz_1;\bar{\aa}_1)\, {\cal Y}(\zz_2,\bar \zz_2;\bar{\aa}_2) \right\rangle =\frac{2\rC_y(\bar{\aa}_1-\bar{\aa}_2)}{(\zz_1-\zz_2)(\bar{\zz}_1-\bar{\zz}_2)^2}\,.
\end{equation}
This operator $\cal Y$ twists the left-moving $SU(2)$ of $SO(4)\simeq SU(2)\times SU(2)$ R-symmetry into the left-moving $SO(2,1)$ of the  
$SO(2,2)\simeq SO(2,1)\times SO(2,1)$ conformal group.

To see that, note that the doublets $a^\alpha$ are in the fundamental of 
the left-moving $SU(2)$. We can implement then 
the R-symmetry transformations of $y^a$ in terms of the following differential operators acting 
on $y(\zz,\bar\zz;\aa,\bar\aa)$
\begin{equation}
\label{su2}
R_\zz^-=\partial_\aa\,,
\qquad
R_\zz^0=\aa\partial_\aa-\tfrac{1}{2}\,,
\qquad
R_\zz^+=-\aa^2\partial_\aa+\aa\,.
\end{equation}
The action of the conformal group on a field is implemented, as usual by
\begin{equation}
\label{so2,1}
P_\zz=\partial_\zz\,,
\qquad
D_\zz=\zz\partial_\zz+\delta\,,
\qquad
K_\zz=\zz^2\partial_\zz+2\zz\delta\, . 
\end{equation}
Note that the dimension $\delta$ for the field $y$ is  equal to 1 (the  total conformal dimension is 
$\Delta= \delta + \bar \delta =2$).

Now we can translate ${\cal Y}(\zz,\bar\zz;\bar{\aa})$ via
\begin{equation}
\partial_\zz{\cal Y}(\zz,\bar\zz;\bar{\aa})=(\partial_\zz+\partial_\aa)y(\zz,\bar\zz;\aa,\bar\aa)\Big|_{\aa=\zz}\,.
\end{equation}
Hence $\cal Y$ transforms covariantly under the twisted generators 
\begin{equation}
\label{twisted-sym}
P_\zz+R_\zz^-\,,\qquad
D_\zz+R_\zz^0\,,\qquad
K_\zz-R_\zz^+\,.
\end{equation}

Let us now examine the supersymmetry constraints on the twisted operators.
The surface defect breaks the $OSp(8^*|4)$ of the ${\cal N}=(2,0)$ 6d theory to 
$[OSp(4^*|2)]^2$. The $OSp(8^*|4)$ transformations can be parameterized
by the spinor $\epsilon(x)=\epsilon^0+\bar\epsilon^1 x^\mu\gamma_\mu$, 
where $\epsilon^0$ corresponds to the super-Poincar\'e transformations and 
$\bar\epsilon^1$ to the superconformal ones. They are 
chiral and antichiral respectively, are in a spinor representation of the $Sp(4)$ R-symmetry 
group and satisfy the symplectic Majorana condition 
$\bar{\epsilon} =-c \Omega \epsilon$, where $c$ is the charge conjugation 
matrix and $\Omega$ the symplectic form.

Using $\gamma_\mu$ and $\rho^a$ for space-time and R-symmetry gamma matrices, 
the surface in the $(x^1,x^2)$ plane ($\zz= x^1 + i x^2$)  imposes the condition
\begin{equation}
\epsilon(x)(1+i\gamma_{12}\rho^5)=0\,.
\end{equation}
Defining  $\gamma_\zz=\ha (\gamma_1-i\gamma_2)$ and likewise $\gamma_{\bar \zz}$, 
we have  $i\gamma_{12}=\gamma_\zz\gamma_{\bar \zz}-\gamma_{\bar \zz}\gamma_\zz$. 
Then the above equation splits into
\begin{equation}
\epsilon_\zz(x)(1+\rho^5)=\epsilon_{\bar \zz}(x)(1-\rho^5)=0\,,
\end{equation}
where $\epsilon_\zz=\epsilon\gamma_\zz$ and $\epsilon_{\bar \zz}=\epsilon\gamma_{\bar \zz}$ 
are the generators of the two $D(2,1;-\frac{1}{2})$ superalgebras.

In addition to the generators in \eqref{su2} and \eqref{so2,1}, each algebra has another 
$SU(2)$ acting on the $x^i$ fields (whose generators we denote by $T$) 
and supercharges $Q$ and $S$. The algebra (see e.g. \cite{Kac:1977qb,Frappat:1996pb}) is
\be
\label{algebra}
\begin{aligned}{}
\{Q_\zz^{\alpha m},Q_\zz^{\beta n}\}&=\varepsilon^{\alpha\beta}\varepsilon^{nm}P_\zz\,,
\\
\{Q_\zz^{\alpha m},S_\zz^{\beta n}\}&=-\varepsilon^{\alpha\beta}\varepsilon^{nm}D_\zz
-2\varepsilon^{mn}(\varepsilon\sigma^i)^{\alpha\beta}R^i_\zz\,
+\varepsilon^{\alpha\beta}(\varepsilon\sigma^i)^{nm}T^i_\zz\,,
\\
\{S_\zz^{\alpha m},S_\zz^{\beta n}\}&=\varepsilon^{\alpha\beta}\varepsilon^{nm}K_\zz\,,
\end{aligned}
\ee
where $\varepsilon=i\sigma^2$.

The supersymmetry transformations of the $y$ fields are
\be
Q_\zz^{\alpha m} y^{\beta\dot\beta}=\varepsilon^{\alpha\beta}\psi^{m\dot\beta}\,.
\ee
For the twisted field at $z\neq0$ we need to also include the $S$ transformations
\be
Q_\zz^{\alpha m} {\cal Y}(\zz,\bar\zz;\bar{\aa})=a^\alpha\psi^{m\dot\beta}\bar a_{\dot\beta}\Big|_{\aa=\zz}\,,
\qquad
S_\zz^{\alpha m} {\cal Y}(\zz,\bar\zz;\bar{\aa})=a^\alpha \zz\psi^{m\dot\beta}\bar a_{\dot\beta}\Big|_{\aa=\zz}\,.
\ee
Thus ${\cal Y}(\zz,\bar\zz;\bar{\aa})$ is annihilated by the combinations 
${\cal Q}_\zz^m=Q_\zz^{2m}-S_\zz^{1m}$.

Let us define a fermionic field which is the action of one of the supersymmetry generators 
on ${\cal Y}$
\be
\Psi^{n}(\zz,\bar\zz;\bar{\aa})
=\frac{1}{\zz}Q_\zz^{2n}{\cal Y}
=\psi^{n\dot\beta}\bar a_{\dot\beta}
=\psi^{n\dot1}+\psi^{n\dot2}\bar{\aa}\,.
\ee
We now examine the Ward identity for the supercharge ${\cal Q}^m_\zz$ for the correlation 
function of this fermion with any number of $\cal Y$ fields. Since ${\cal Q}_\zz^m$ annihilates 
$\cal Y$, we get
\be
\label{ward}
\begin{aligned}
0&=\big\langle{\cal Q}^m[\Psi^n(\zz_1,\bar\zz_1;\bar{\aa}_1){\cal Y}(\zz_2,\bar\zz_2;\bar{\aa}_2)\cdots{\cal Y}(\zz_p,\bar\zz_p;\bar{\aa}_p)]\big\rangle
=\big\langle[{\cal Q}^m\Psi^n(\zz_1,\bar\zz_1;\bar{\aa}_1)]\,{\cal Y}(\zz_2,\bar\zz_2;\bar{\aa}_2)\cdots{\cal Y}(\zz_p,\bar\zz_p;\bar{\aa}_p)\big\rangle
\\
&=\frac{1}{\zz_1}\big\langle[\{{\cal Q}_\zz^m,Q_\zz^{2n}\}{\cal Y}(\zz_1,\bar\zz_1;\bar{\aa}_1)]\,
{\cal Y}(\zz_2,\bar\zz_2;\bar{\aa}_2)\cdots{\cal Y}(\zz_p,\bar\zz_p;\bar{\aa}_p)\big\rangle\,,
\end{aligned}
\ee
where in the last step we have used again that ${\cal Q}_\zz^m$ annihilates ${\cal Y}$, leaving the anticommutator of ${\cal Q}_\zz^m$ and $Q_\zz^{2n}$. Using \eqref{algebra}, this anticommutator evaluates to
\be
\{{\cal Q}_\zz^m,Q_\zz^{2n}\}
=\varepsilon^{nm}D_\zz+2\varepsilon^{nm}R_\zz^0\,,
\ee
and its action on ${\cal Y}(\zz,\bar\zz;\bar{\aa})$ can be expressed as the differential operators \eqref{su2},\eqref{so2,1}
\be
\label{466}
\left(D_\zz+2R_\zz^0\right){\cal Y}(\zz,\bar\zz;\bar{\aa})
=(\zz\partial_\zz+2\aa\partial_\aa)y(\zz,\bar\zz,\aa,\bar\aa)\big|_{\aa=\zz}\,.
\ee

Let us apply this relation to the 4-point function \rf{777},\rf{430} of the ${\cal Y}(\zz,\bar \zz;\bar{\aa})$ operators.
Examining the definitions of  conformal cross-ratios in \eqref{44}, we conclude that in this case 
$\alpha=1/\ze$, i.e. 
\be \la{445}
\begin{split}
G_{\text{left-twist}}(\zz_i,\bar{\zz}_i;\bar{\aa}_i)&=\langle{\cal Y}(\zz_1,\bar\zz_1;\bar{\aa}_1) {\cal Y}(\zz_2,\bar\zz_2;\bar{\aa}_2) {\cal Y}(\zz_3,\bar\zz_3;\bar{\aa}_3) {\cal Y}(\zz_4,\bar\zz_4;\bar{\aa}_4) \rangle\\
{}&={\frac{4|\aa_1-\aa_2|^2|\aa_3-\aa_4|^2}{|\zz_1-\zz_2|^4|\zz_3-\zz_4|^4}}{\cal G} ( \ze, \bar \ze; \alpha,\bar{\alpha})\Big{|}_{\aa_i=\zz_i}\\
{}&= {\frac{4(\bar{\aa}_1-\bar{\aa}_2)(\bar{\aa}_3-\bar{\aa}_4)}{(\zz_1-\zz_2)\, (\zz_3-\zz_4)(\bar{\zz}_1-\bar{\zz}_2)^2\, (\bar{\zz}_3-\bar{\zz}_4)^2}}{\cal G} ( \ze, \bar \ze; 1/\ze,\bar{\alpha}) \,. 
\end{split}
\ee
Noting that the differential operator \eqref{466} annihilates the prefactor in the second line above, we can translate the Ward identity \eqref{ward}, \eqref{466} to act on the function of cross ratios ${\cal G}$. 
Moreover, if we 
fix $\zz_1=0$, $\zz_3=1$ and $\zz_4\to\infty$, we  find from \rf{44}  that $\ze=\zz_2$, and since the 
twisting \rf{4522}
fixes $\aa_2=\zz_2$, then $\alpha=1/\aa_2$, and we reproduce precisely the first relation in \eqref{scfwardid}.

Note that our construction has only used the left-moving copy of $D(2,1;-\frac{1}{2})$. We can repeat the entire analysis by twisting instead  the right-moving $D(2,1;-\frac{1}{2})$, which leads to the second identity in (\ref{scfwardid}). Thanks to the factorization property, our derivation above implies that the superconformal Ward identity
\begin{equation}
\Big({-\frac{1}{2}}\ze\partial_\ze+\alpha\partial_\alpha\Big)\mathcal{F}(\ze;\alpha)\Big|_{\alpha=1/\ze}=0\;,
\end{equation}
applies as well to the 4-point functions of $1\ov  2$-BPS insertions on a line defect in 4d $\mathcal{N}=2$ theories preserving half of the supersymmetry. This system has the same $D(2,1;-\frac{1}{2})=OSp(4^*|2)$ symmetry \cite{Gimenez-Grau:2019hez}. Here $\ze$ is the conformal cross ratio on a straight line, and $\alpha$ is the cross ratio for the $SO(3)$ R-symmetry.

Although we derived (\ref{scfwardid}) for correlators of fields with conformal dimensions 
$(\delta,\bar{\delta})=(1,1)$ and R-symmetry charges $(q,\bar{q})=(\frac{1}{2},\frac{1}{2})$, 
the same superconformal Ward identities (\ref{scfwardid}) apply to 4-point correlators of 
dCFT operators with general R-symmetry charges $(q,\bar{q})$ and $(\delta,\bar{\delta})=(2q,2\bar{q})$. 
That $q$ and $\bar{q}$ do not need to be  equal is a consequence of the factorized form of the 
superconformal algebra $[D(2,1;-\frac{1}{2})]^2$. We can understand the extension 
to general $1\ov  2$-BPS insertions by realizing that we can construct higher-weight $1\ov  2$-BPS operators by 
taking products of the ones with $\delta=2q=1$ or $\bar{\delta}=2\bar{q}=1$. The $n$-point correlators 
of operators with lowest weights satisfy the constraint (\ref{ward}), which becomes (\ref{scfwardid}) 
when regrouping them into four composite operators.

\subsection{Twisted 4-point correlator and a curious relation to  special 4-point function  in $\N=4$ SYM 
 } 
 \label{sec:twisted4pt}

We can twist both $D(2,1;-\frac{1}{2})$  algebras with $\aa=\zz$ and $\bar{\aa}=\bar{\zz}$. This gives a dimension-one non-chiral scalar operator 
\begin{equation}
\left\langle y(\zz_1,\bar \zz_1;\zz_1,\bar{\zz}_1)\, y(\zz_2,\bar \zz_2;\zz_2,\bar{\zz}_2) \right\rangle =\frac{2\rC_y}{|\zz_1-\zz_2|^2}\,.
\end{equation}
Now take the 4-point function of such double-twisted operators
\begin{equation}
G_{\rm twist}(\zz_i,\bar{\zz}_i;\aa_i,\bar{\aa}_i)=G(\zz_i,\bar{\zz}_i;\zz_i,\bar{\zz}_i)\;.
\end{equation}
Using \rf{488},\rf{49}  we find that 
the correlator   \rf{445}  has a  remarkably  simple expression proportional to just one  $D$-function
\begin{equation}\la{450}
G_{\rm twist}=-\frac{96 N}{\pi^4 }\vec{x}_{12}^2\vec{x}_{23}^2\vec{x}_{24}^2D_{2422}\;.
\end{equation}
Surprisingly, the same function arises in a totally different setting, 
namely, in the 4-point function of the stress tensor multiplet of $\N=4$ SYM theory at strong coupling computed 
from the AdS$_5\times S^5$ IIB supergravity. 

Indeed, let us  consider the 4-point function of the $1\ov  2$-BPS operator $\mathcal{O}_2(\vec{x}; {\rm t})={\rm t}_I {\rm t}_J\,  \tr (\Phi^I\Phi^J)(\vec{x}) $  where 
$\Phi^I$  ($I=1,\cdots, 6$) are the 6 real  scalars of $\mathcal{N}=4$ SYM.\footnote{We will be brief in the following about the superconformal kinematics of $1\ov  2$-BPS 4-point functions in $\mathcal{N}=4$ SYM, and refer the interested reader to Section~2 of \cite{Alday:2019nin} for a more detailed review.} This  operator has protected conformal dimension $\Delta=2$, and transforms in the rank-2 symmetric traceless representation of $SO(6)$ R-symmetry. We contracted the indices with a null vector ${\rm t}_I$ satisfying ${\rm t}^2=0$, which automatically performs the projection to the symmetric traceless representation. Thanks to superconformal symmetry, the 4-point function has a ``partially non-renormalized'' structure \cite{Eden:2000bk,Nirschl:2004pa}
\begin{equation}
G_{\rm SYM}(\vec{x}_i;{\rm t}_i)=\langle \mathcal{O}_2(\vec{x}_1;{\rm t}_1)\mathcal{O}_2(\vec{x}_2;{\rm t}_2)\mathcal{O}_2(\vec{x}_3;{\rm t}_3)\mathcal{O}_2(\vec{x}_4;{\rm t}_4) \rangle\;, 
\end{equation}
\begin{equation}
G_{\rm SYM}(\vec{x}_i;{\rm t}_i)=G_{\rm free}(\vec{x}_i;{\rm t}_i)+ R(\vec{x}_i;{\rm t}_i)   \,H(\vec{x}_i)\;. \la{452}
\end{equation}
Here $G_{\rm free}(\vec{x}_i;{\rm t}_i)$ is the correlator in the free SYM  theory
\begin{equation}\la{461}
G_{\rm free}(\vec{x}_i;{\rm t}_i)=\frac{{\rm t}_{12}^2{\rm t}_{34}^2}{\vec{x}_{12}^4\vec{x}_{34}^4}\Big[\Big(1+\sigma^2U^2+\tau^2\frac{U^2}{V^2}\Big)+\frac{1}{c}\Big(\sigma U+\tau \frac{U}{V}+\sigma\tau \frac{U^2}{V}\Big)\Big]\;, 
\end{equation}
where we assumed  the canonical normalization $ \langle \mathcal{O}_2(\vec{x}_1;{\rm t}_1)\mathcal{O}_2(\vec{x}_2;{\rm t}_2) \rangle=
\frac{{\rm t}_{12}^2}{\vec{x}_{12}^4}$.
Note that  the free correlator is exact in $1/c$, where $c=\frac{1}{4} (N^2-1)$ is the ``central charge" of the $SU(N)$ SYM theory.
$R$ in  \rf{452}   is a  kinematical factor fully determined by the superconformal symmetry
\begin{equation}
R={\rm t}_{12}^2{\rm t}_{34}^2\, \vec{x}_{13}^4\vec{x}_{24}^4(1-\ze\alpha)(1-\ze\bar{\alpha})(1-\bar{\ze}\alpha)(1-\bar{\ze}\bar{\alpha})\;,
\end{equation}
and $H(\vec{x}_i)$ is the {\it reduced correlator} which encodes all the dynamical information. 
We can compute $H(\vec{x}_i)$ in $1/c$ expansion at strong coupling, using the dual bulk description of IIB supergravity on AdS$_5\times S^5$
\begin{equation}
H(\vec{x}_i)=H_{\rm tree}(\vec{x}_i)+H_{\text{1-loop}}(\vec{x}_i)+\ldots\;.
\end{equation}
The tree-level reduced correlator reads \cite{Arutyunov:2000py}
\begin{equation}\la{456}
H_{\rm tree}(\vec{x}_i)=-\frac{6}{\pi^2 c}\frac{D_{2422}}{\vec{x}_{13}^2\vec{x}_{34}^2\vec{x}_{14}^2}\;.
\end{equation}
Comparing this  to  the twisted correlator in \rf{450} \footnote{The reader might be concerned that we are comparing results in different spacetime dimensions. However, a nice feature of $D$-functions defined for $AdS_{d+1}$ is that the $d$-dependence only appears in the overall normalization. The functional dependence on $\vec{x}_{ij}^2$ is the same for all $d$. Moreover, for four points we can always use a conformal transformation to restrict them on a two-dimensional plane.} we conclude that they match up to an overall constant   and 
a factor that   can be interpreted as 
a ``tetrahedron'' contraction of generalized free fields\footnote{Here we do not want to absorb $\mathcal{T}$ into the definition of the reduced correlator $H_{\rm tree}$, because it is important that $H_{\rm tree}$ has conformal dimension 4 to exhibit the ten-dimensional hidden conformal symmetry \cite{Caron-Huot:2018kta}. Replacing the argument $\vec{x}_{ij}^2$ in $H_{\rm tree}(\vec{x}_{ij}^2)$ with $\vec{x}_{ij}^2+{\rm t}_{ij}$ gives a generating function for the reduced correlators of higher Kaluza-Klein modes (see Section~2.3 of~\cite{Alday:2019nin} for a discussion of this point).}
\begin{equation}
\mathcal{T}=\prod_{1\leq i<j\leq 4}{1\ov \vec{x}_{ij}^2}\;.
\end{equation}
The observed relation\footnote{It might also be instructive to view the relation from the Mellin perspective. For the twisted correlator \rf{450}, we have a factorized polynomial Mellin amplitude 
\begin{equation}
\no \mathcal{M}_{\rm twisted}\sim (s-2)(t-2)(u-2)\;, \quad s+t+u=4\;,
\end{equation}
and it has an interpretation of contact interactions with up to six derivatives. For the supergravity case \rf{456}, the Mellin amplitude is the  inverse of the  above expression
\begin{equation}
\no \mathcal{M}_{\rm supergravity}\sim \frac{1}{(s-2)(t-2)(u-2)}\;, \quad s+t+u=4\;.
\end{equation}
} between the twisted 
correlator on BPS surface defect in 6d $(2,0)$ theory, and  the BPS   correlator in strongly coupled $\mathcal{N}=4$ SYM  theory 
is quite curious and we hope  to shed light on its meaning in the future.

\subsection{Fixing the $\langle yyyy \rangle$ correlator  from  the superconformal  Ward identities} 
\label{sec:ward}

In this subsubsection, we provide an alternative perspective on the tree-level holographic 4-point function $\langle yyyy \rangle$. We show that the holographic correlator \rf{4.7}  can be ``bootstrapped'' by imposing superconformal constraints captured by the the relations (\ref{scfwardid}), without using any precise information about the  coefficients of the bulk vertices. 
Similar techniques have already been implemented in a number of maximally supersymmetric AdS backgrounds, and lead to unique answers for tree level 4-point functions in theories with no defects \cite{Rastelli:2016nze,Rastelli:2017udc,Rastelli:2017ymc,Zhou:2017zaw}.

We start from an ansatz  for a local \adst bulk   action which consists of  all possible 
contact interactions  of $y^a$ fields    with up to four derivatives.  The  structure of the  corresponding   Witten diagrams 
 translates 
into the ansatz for the  following  tree-level 4-point function 
\begin{equation}
G=G_1+G_2\ , \la{418}
\end{equation}
where  $G_1$ is  the parity-even part   given  by  a linear combination of
contributions of  all possible 0-, 2-, and 4-derivative contact diagrams
and   $G_2$  is  the parity odd part  coming from   the 3-derivative contact interaction. The precise combination of these Witten diagrams can be fixed as in Section \ref{sec:yyyy} using the  explicit  form of the  M2 brane  Lagrangian (\ref{39}), but here we will leave them arbitrary and to be determined  by symmetries. 

A convenient parameterization for $G_1$ is given by the linear combination of all possible $D$-functions that can show up at this order  with the  coefficients that 
are  any  possible parity-even R-symmetry structures
\begin{align}\la{4190}
&G_1 = \sum_i      (  \mu_{1,i}\,  \tt_{12}\tt_{34}+\mu_{2,i} \, \tt_{13}\tt_{24}+\mu_{3,i}\,  \tt_{14}\tt_{23})\, W_i \ , \\
&\{W_i\}=\{D_{2222};\ 
\vx_{12}^2 D_{3322},\ \vx_{13}^2 D_{3232}, \cdots\;; \   
\vx_{12}^2\vx_{34}^2 D_{3333},\ \vx_{13}^2\vx_{24}^2D_{3333},  \cdots\;\} \ . \la{4199}
\end{align}
We also require that the parity-even part  $G_1$  (as coming  from a local bulk action)   should be crossing symmetric. 

The  only    parity-odd   4-vertex allowed  by symmetries  is the one in \rf{40}, i.e. 
 $\sim \ep_{abcd} \ep^{\m\n\l} y^a \del_\m y^b \del_\n y^c \del_\l y^d$ 
   and thus  $G_2$ is proportional to the 3-derivative contact Witten diagram in AdS$_3$  
discussed  in \cite{Rastelli:2019gtj} (cf. \rf{111}) 
\begin{equation}\la{414}
G_2 \sim  \int \frac{d^3x}{\xxx ^3} \xxx ^3\, \epsilon^{\mu\nu\rho} \partial_\mu G^{\Delta_1}_{B\partial}(x,\vx_1)\partial_\nu G^{\Delta_3}_{B\partial}(x,\vx_3)\partial_\rho G^{\Delta_4}_{B\partial}(x,\vx_4) G^{\Delta_2}_{B\partial}(x,\vx_2)\;.
\end{equation}
It is easy to check by integration by parts that this contribution 
is antisymmetric with respect to all four points $\vx_i$. In order for $G_2$ to be crossing symmetric, the R-symmetry factor has to be anti-symmetric and can only be 
\begin{equation}
\tt_{12}\tt_{34}(\alpha-\bar{\alpha})\;.
\end{equation}
Using the result of \cite{Rastelli:2019gtj}, $G_2$ can be written as (cf. \rf{49})
\begin{equation}
G_2=\lambda\, \frac{\tt_{12}\tt_{34}}{\vx_{12}^4\vx_{34}^4}U^2 (\ze-\bar{\ze})(\alpha-\bar{\alpha})\bar{D}_{3333}\ , 
\end{equation}
where $\lambda$ is an undetermined coefficient.

Imposing the superconformal Ward identities (\ref{scfwardid}), we find that all the coefficients $\mu_k$ 
in the ansatz \rf{418} can be fixed except for an overall scaling factor.\footnote{To implement the superconformal Ward identities, we used the algorithm developed in \cite{Rastelli:2017udc} (see Section~5 of the reference for notation and details). We decompose all the $\bar{D}$-functions into the basis spanned by 1, $\log U$, $\log V$ and the 1-loop scalar box diagram $\Phi(U,V)$, by using differential recursion relations of $\Phi(U,V)$. The superconformal Ward identities are expanded into this basis, with rational functions as coefficient functions. Requiring the coefficient functions to vanish gives linear equations for the unfixed coefficients in the ansatz.}   
The overall normalization is not  fixed  by    (\ref{scfwardid})  because 
these relations  are homogenous. The solution is proportional to $\mathcal{G}_{\rm tree}$ we obtained 
above  \rf{4.7} \rf{488},\rf{49} by the direct computation from the M2-brane action.

Thus the superconformal symmetry is effectively determining the  relative  coefficients in  the underlying 
bulk action. Note that even though we included the zero-derivative contact interactions in the ansatz for $G_1$ in \rf{4190}, such contributions are absent in the final solution fixed by superconformal symmetry. 
This is consistent with the
fact that the  are no such terms in the M2-brane action  for $y^a$ in \rf{39}.

We can also apply  a similar bootstrap approach  to  the   case of 
$1\ov  2$-BPS Wilson loop. The corresponding  4-point function
for the dimension 1 operators 
can be uniquely fixed by the superconformal Ward identity, 
up to an overall constant which can be determined
 using supersymmetric localization \cite{Giombi:2017cqn}. 
 We will give the details of this calculation in Appendix~\ref{sec:WL}.

\section*{Acknowledgments}
We are indebted to the CERN theory group for hosting us at the 
``Exact Computations in AdS/CFT'' workshop, where 
seeds of this work were sown. 
The work of N.D. is supported by an STFC grant number  ST/P000258/1. He is grateful to 
M. Probst and M. Tr\'epanier for stimulating discussions and related collaboration and 
the hospitality of EPFL Lausanne. 
The work of S.G. is supported in part by the US NSF under Grants No.~PHY-1620542 and PHY-1914860.
A.T.  thanks  M. Beccaria and R. Roiban for useful  discussions  and   acknowledges the support of the 
 STFC grant ST/P000762/1. 
  The work of X.Z. is supported in part by the Simons Foundation Grant No. 488653.
We also  thank M. Meineri and L. Bianchi for comments on the first version of this paper.

\appendix

\section{$\bar{D}$-functions}
For reader's convenience, we collect some useful properties of the $\bar{D}$-functions  in \rf{4160},\rf{4170}) 
 (see, e.g., \cite{Arutyunov:2002fh}), which can be used to obtain the explicit form of correlators as functions of cross ratios
 $U$ and $V$ or $\ze$ and $\bar \ze$  in \rf{44}.

The simplest $\bar{D}$-function has $\Delta_i=1$, and is just the scalar one-loop box digram in four dimensions
\begin{align}
\bar{D}_{1111}&=\Phi(\ze,\bar{\ze})\;, \\
\Phi(\ze,\bar{\ze})&=\frac{1}{\ze-\bar{\ze}}\Big[\log(\ze\bar{\ze})\log\Big(\frac{1-\ze}{1-\bar{\ze}}\Big)+2{\rm Li}_2(\ze)-2\rm{Li}_2(\bar{\ze})\Big]\;.
\end{align}
To obtain $\bar{D}$-functions with higher weights, we can use the following differential operators 
\begin{equation}
\begin{split}
\bar{D}_{\Delta_1+1,\Delta_2+1,\Delta_3,\Delta_4}&=-\partial_U \bar{D}_{\Delta_1,\Delta_2,\Delta_3,\Delta_4}\;,\\
\bar{D}_{\Delta_1,\Delta_2,\Delta_3+1,\Delta_4+1}&=(\Delta_3+\Delta_4-\Sigma-U\partial_U )\bar{D}_{\Delta_1,\Delta_2,\Delta_3,\Delta_4}\;,\\
\bar{D}_{\Delta_1,\Delta_2+1,\Delta_3+1,\Delta_4}&=-\partial_V \bar{D}_{\Delta_1,\Delta_2,\Delta_3,\Delta_4}\;,\\
\bar{D}_{\Delta_1+1,\Delta_2,\Delta_3,\Delta_4+1}&=(\Delta_1+\Delta_4-\Sigma-V\partial_V )\bar{D}_{\Delta_1,\Delta_2,\Delta_3,\Delta_4}\;,\\
\bar{D}_{\Delta_1,\Delta_2+1,\Delta_3,\Delta_4+1}&=(\Delta_2+U\partial_U+V\partial_V )\bar{D}_{\Delta_1,\Delta_2,\Delta_3,\Delta_4}\;,\\
\bar{D}_{\Delta_1+1,\Delta_2,\Delta_3+1,\Delta_4}&=(\Sigma-\Delta_4+U\partial_U+V\partial_V )\bar{D}_{\Delta_1,\Delta_2,\Delta_3,\Delta_4}
\end{split}
\end{equation}
where $\Sigma=\frac{1}{2}(\Delta_1+\Delta_2+\Delta_3+\Delta_4)$. 

Note that  the function $\Phi(\ze,\bar{\ze})$ satisfies the following differential recursion relations
\begin{equation}
\begin{split}
\partial_\ze\Phi&=-\frac{1}{\ze-\bar{\ze}}\Phi-\frac{1}{\ze(\ze-\bar{\ze})}\log(1-\ze)(1-\bar{\ze})+\frac{1}{(-1+\ze)(\ze-\bar{\ze})}\log(\ze\bar{\ze})\;,\\
\partial_{\bar{\ze}}\Phi&=\frac{1}{\ze-\bar{\ze}}\Phi+\frac{1}{\bar{\ze}(\ze-\bar{\ze})}\log(1-\ze)(1-\bar{\ze})-\frac{1}{(-1+\bar{\ze})(\ze-\bar{\ze})}\log(\ze\bar{\ze})\;.
\end{split}
\end{equation}
We can therefore recursively decompose $\bar{D}_{\Delta_1,\Delta_2,\Delta_3,\Delta_4}$ into a basis spanned by 1, $\log U$, $\log V$, $\Phi(\ze,\bar{\ze})$  with   coefficients   being  rational functions of $\ze$, $\bar{\ze}$.

\section{4-point 
correlator  on  $1\ov  2$-BPS Wilson line 
 from\\ superconformal invariance}
\label{sec:WL}

In this Appendix  we implement the techniques of Section~\ref{sec:ward} to determine the tree-level contribution 
to the the 4-point function of 
insertions in the $1\ov  2$-BPS Wilson loop from the superconformal Ward identity and crossing. 
This reproduces the expression derived in~\ci{Giombi:2017cqn} from the 
fundamental string action in AdS$_5\times S^5$. 

Recall that the Wilson loop has an $OSp(4^*|4)$ superconformal symmetry. The insertions 
$\Phi^a$ with $a=1,\cdots, 5$ have conformal dimension 1, and transforms as a vector under the 
$SO(5)=Sp(4)$ R-symmetry.  Holographically, they corresponds to the $S^5$ fluctuation $y^a$.
As in \rf{420}, we contract the R-symmetry index of $\Phi^a$ with a constant auxiliary vector ${\rm t}_a $ 
\begin{equation}
\Phi(w;{\rm t})={\rm t}_a\, \Phi^a(w)\;,
\end{equation}
so that its correlators depend  on the coordinates $w_i \in \mathbb{R}^1 $  parametrizing the straight Wilson line 
and on the internal ``coordinates" ${\rm t}_i$. 

For simplicity, we will fix the normalization  of the two-point function  so that 
\begin{equation}
\llangle \Phi(w_1;{\rm t}_1)\Phi(w_2; {\rm t}_2)\rrangle=\frac{{\rm t}_{12}}{|w_{12}|^2}\ , 
\end{equation}
where ${\rm t}_{ij}={\rm t}_{i}\cdot {\rm t}_{j}$, $w_{ij}=w_i-w_j$. The 4-point function can be written 
in terms of  a function $\mathcal{F}$ of cross ratios as 
\begin{equation}\la{a3} 
A= \llangle \Phi(w_1; {\rm t}_1)\Phi(w_2; {\rm t}_2)\Phi(w_3; {\rm t}_3)\Phi(w_4; {\rm t}_4)\rrangle=\frac{{\rm t}_{12}{\rm t}_{34}}{|w_{12}|^2|w_{34}|^2}\, \mathcal{F}(\ze;\alpha,\bar{\alpha})\ ,
\end{equation}
where  
\begin{equation}
\ze=\frac{w_{12}w_{34}}{w_{13}w_{24}}\;,\qquad \sigma=\frac{{\rm t}_{13}{\rm t}_{24}}{{\rm t}_{12}{\rm t}_{34}}=\alpha\bar{\alpha}\;,
\qquad \tau=\frac{{\rm t}_{14}{\rm t}_{23}}{{\rm t}_{12}{\rm t}_{34}}=(1-\alpha)(1-\bar{\alpha})\;.
\end{equation}
Note that unlike the  $SO(4)$ case of the M2-brane theory in the main text,  here 
the R-symmetry cross ratios $\alpha$, $\bar{\alpha}$  should 
appear symmetrically in $\mathcal{F}(\ze;\alpha,\bar{\alpha})$ (we cannot have $\det({\rm t}_{ij})\propto {\rm t}_{12}{\rm t}_{34}(\alpha-\bar{\alpha})$). In other words, $\mathcal{F}(\ze;\alpha,\bar{\alpha})$ is a linear function of $\sigma$ and $\tau$. The 4-point function also needs to satisfy the superconformal Ward identities  \cite{Liendo:2018ukf}  analogous to 
\rf{scfwardid}
\begin{equation}\label{WLscfwardid}
\Big({-\frac{1}{2}}\ze\partial_\ze+\alpha\partial_\alpha\Big)\mathcal{F}(\ze;\alpha,\bar{\alpha})\bigg|_{\alpha=1/\ze}=0\;, \qquad 
\Big({-\frac{1}{2}}\ze\partial_\ze+\bar{\alpha}\partial_{\bar{\alpha}}\Big)\mathcal{F}(\ze;\alpha,\bar{\alpha})\bigg|_{\bar{\alpha}=1/\ze}=0\;.
\end{equation}
Because $\mathcal{F}(\ze;\alpha,\bar{\alpha})$ is symmetric under $\alpha\leftrightarrow\bar{\alpha}$, the second equation is redundant. Moreover, if we set $\alpha=\bar{\alpha}=1/\ze$, (\ref{WLscfwardid}) implies that the twisted correlator 
(the analog of \rf{450}) is now topological
\begin{equation}\label{topotwisted}
\partial_\ze \mathcal{F}(\ze;1/\ze,1/\ze)=0\;.
\end{equation}

We take an ansatz for the tree-level contribution to  the correlator $A$ in \rf{a3}
which is essentially the same as for the parity-even part $G_1$ in \rf{418},\rf{4190}
in Section~\ref{sec:ward}. 
It includes all $D$-functions that can appear in  AdS$_2$ 
contact  Witten diagrams with   up to four derivative  vertices 
\begin{equation}
\begin{split}  
\{W_i\}=\{D_{1111}; \  w_{12}^2 D_{2211},\  w_{13}^2 D_{2121}, \cdots;\  w_{12}^2w_{34}^2 D_{2222},\  w_{13}^2w_{24}^2D_{2222}\;, \cdots\;\}\ . 
\end{split}
\end{equation}
The ansatz for $A$, the analog of~\rf{4190}, now has only coefficients that are linear in $\sigma$ and $\tau$
\begin{equation}\la{a8}
A={\rm t}_{12}{\rm t}_{34}\sum_i (\mu_{1,i}+\mu_{2,i}\,  \sigma+\mu_{3,i}\, \tau)\, W_i \ . 
\end{equation}
Note that the  dimension $d=p$ dependence of  the $D$-functions
only comes as an overall factor and does not affect the dependence on 
the cross ratios (as is clear, for example, from the Mellin representation). 
Therefore, we can compute them for  generic $d$, and set $\ze=\bar{\ze}$, $d=1$ in the end. 

We now require that  $A$ should be  (i) crossing symmetric, and (ii) 
satisfy the superconformal Ward identity (\ref{WLscfwardid}).  Remarkably, 
this allows us  to determine all  the coefficients $\mu_{r,i}$ in the ansatz \rf{a8}   
up to an overall factor $\nu$  (cf. \rf{a3}) 
\begin{align}
A&=\nu\,  \frac{{\rm t}_{12}{\rm t}_{34}}{|w_{12}|^2|w_{34}|^2}\mathcal{A}\;, \qquad \  \mathcal{F} = \nu \mathcal{A}\ , \la{a9}
\\
\la{a10}
\mathcal{A}(\ze;\alpha,\bar{\alpha})&=
\frac{\ze^2-2 \ze+2}{\ze-1}-\sigma \frac{  \ze \left(2 \ze^2-2 \ze+1\right)}{\ze-1}-\tau \frac{  \ze \left(\ze^2+1\right)}{(\ze-1)^2}
\nonumber
\\
{}&+\Big[ \frac{(\ze-1) \left(\ze^2+\ze+2\right)}{\ze}-\sigma  (\ze-1) \left(2 \ze^2+\ze+1\right)-\tau  \left(\ze^2+1\right)\Big] \log(1-\ze)\\
{}&+\Big[{-}\frac{\left(\ze^2-2 \ze+2\right) \ze^2}{(\ze-1)^2}+\sigma \frac{  \left(2 \ze^2-5 \ze+4\right) \ze^3}{(\ze-1)^2}+
\tau  \frac{ \left(\ze^2-3 \ze+4\right) \ze^3}{(\ze-1)^3}\Big]\log \ze\;.
\nonumber
\end{align}
Note that as predicted in (\ref{topotwisted}) the twisted correlator is a constant  since 
\begin{equation}
\mathcal{F}(\ze;1/\ze,1/\ze) = \nu \mathcal{A}(\ze;1/\ze,1/\ze)=-3\nu \;.
\end{equation}
This twisted correlator, however, can be independently computed using supersymmetric localization 
\cite{Giombi:2017cqn},\footnote{As a side remark, note that the Ward identities (\ref{a10}) and the topological condition (\ref{topotwisted}) also apply to the general correlation functions of $1\ov  2$-BPS insertions with arbitrary charges. The resulting topological correlators can also be computed exactly from localization \cite{Giombi:2018qox, Liendo:2018ukf, Giombi:2018hsx}.} so that at leading order in the inverse string tension we should have 
\begin{equation}
\mathcal{F}(\ze;1/\ze,1/\ze)=-\frac{3}{\sqrt{\lambda}}+\mathcal{O}(\frac{1}{\lambda})\;.
\end{equation}
This fixes the overall factor to be 
\begin{equation}
\nu=\frac{1}{\sqrt{\lambda}}\;.
\end{equation}
The resulting expression \rf{a9},\rf{a10} agrees with 
the one  found in \ci{Giombi:2017cqn} directy from the string action.


\bibliographystyle{utphys2}
\bibliography{M2-bib}

\providecommand{\href}[2]{#2}\begingroup\raggedright\begin{thebibliography}{10}\setlength{\parskip}{1pt}\setlength{\itemsep}{0pt
  plus 0.3ex}

\bibitem{Maldacena:1998im}
J.~M. Maldacena, ``{Wilson loops in large $N$ field theories},''
  \href{http://dx.doi.org/10.1103/PhysRevLett.80.4859}{{\em Phys. Rev. Lett.}
  {\bfseries 80} (1998) 4859--4862},
\href{http://arxiv.org/abs/hep-th/9803002}{{\ttfamily hep-th/9803002}}.

\bibitem{Berenstein:1998ij}
D.~E. Berenstein, R.~Corrado, W.~Fischler, and J.~M. Maldacena, ``{The operator
  product expansion for Wilson loops and surfaces in the large $N$ limit},''
  \href{http://dx.doi.org/10.1103/PhysRevD.59.105023}{{\em Phys. Rev.}
  {\bfseries D59} (1999) 105023},
\href{http://arxiv.org/abs/hep-th/9809188}{{\ttfamily hep-th/9809188}}.

\bibitem{Drukker:2000ep}
N.~Drukker, D.~J. Gross, and A.~A. Tseytlin, ``{Green-Schwarz string in
  AdS$_5\times S^5$: Semiclassical partition function},''
  \href{http://dx.doi.org/10.1088/1126-6708/2000/04/021}{{\em JHEP} {\bfseries
  04} (2000) 021},
\href{http://arxiv.org/abs/hep-th/0001204}{{\ttfamily hep-th/0001204}}.

\bibitem{Cooke:2017qgm}
M.~Cooke, A.~Dekel, and N.~Drukker, ``{The Wilson loop CFT: Insertion
  dimensions and structure constants from wavy lines},''
  \href{http://dx.doi.org/10.1088/1751-8121/aa7db4}{{\em J. Phys.} {\bfseries
  A50} no.~33, (2017) 335401},
\href{http://arxiv.org/abs/1703.03812}{{\ttfamily arXiv:1703.03812}}.

\bibitem{Giombi:2017cqn}
S.~Giombi, R.~Roiban, and A.~A. Tseytlin, ``{Half-BPS Wilson loop and
  AdS$_2$/CFT$_1$},''
  \href{http://dx.doi.org/10.1016/j.nuclphysb.2017.07.004}{{\em Nucl. Phys.}
  {\bfseries B922} (2017) 499--527},
\href{http://arxiv.org/abs/1706.00756}{{\ttfamily arXiv:1706.00756}}.

\bibitem{Constable:2002xt}
N.~R. Constable, J.~Erdmenger, Z.~Guralnik, and I.~Kirsch, ``{Intersecting
  D3-branes and holography},''
  \href{http://dx.doi.org/10.1103/PhysRevD.68.106007}{{\em Phys. Rev.}
  {\bfseries D68} (2003) 106007},
\href{http://arxiv.org/abs/hep-th/0211222}{{\ttfamily hep-th/0211222}}.

\bibitem{Drukker:2008wr}
N.~Drukker, J.~Gomis, and S.~Matsuura, ``{Probing ${\cal N}=4$ SYM with surface
  operators},'' \href{http://dx.doi.org/10.1088/1126-6708/2008/10/048}{{\em
  JHEP} {\bfseries 10} (2008) 048},
\href{http://arxiv.org/abs/0805.4199}{{\ttfamily arXiv:0805.4199}}.

\bibitem{Drukker:2020dcz}
N.~Drukker, M.~Probst, and M.~Tr\'epanier, ``{Surface operators in the 6d N =
  (2, 0) theory},'' \href{http://dx.doi.org/10.1088/1751-8121/aba1b7}{{\em J.
  Phys. A} {\bfseries 53} no.~36, (2020) 365401},
  \href{http://arxiv.org/abs/2003.12372}{{\ttfamily arXiv:2003.12372
  [hep-th]}}.

\bibitem{Grabner:2020nis}
D.~Grabner, N.~Gromov, and J.~Julius, ``{Excited States of One-Dimensional
  Defect CFTs from the Quantum Spectral Curve},''
  \href{http://dx.doi.org/10.1007/JHEP07(2020)042}{{\em JHEP} {\bfseries 07}
  (2020) 042}, \href{http://arxiv.org/abs/2001.11039}{{\ttfamily
  arXiv:2001.11039 [hep-th]}}.

\bibitem{Gustavsson:2004gj}
A.~Gustavsson, ``{Conformal anomaly of Wilson surface observables: A field
  theoretical computation},''
  \href{http://dx.doi.org/10.1088/1126-6708/2004/07/074}{{\em JHEP} {\bfseries
  07} (2004) 074},
\href{http://arxiv.org/abs/hep-th/0404150}{{\ttfamily hep-th/0404150}}.

\bibitem{Ganor:1996nf}
O.~J. Ganor, ``{Six-dimensional tensionless strings in the large $N$ limit},''
  \href{http://dx.doi.org/10.1016/S0550-3213(96)00702-X}{{\em Nucl. Phys.}
  {\bfseries B489} (1997) 95--121},
\href{http://arxiv.org/abs/hep-th/9605201}{{\ttfamily hep-th/9605201}}.

\bibitem{Strominger:1995ac}
A.~Strominger, ``{Open $p$-branes},''
  \href{http://dx.doi.org/10.1016/0370-2693(96)00712-5}{{\em Phys. Lett.}
  {\bfseries B383} (1996) 44--47},
\href{http://arxiv.org/abs/hep-th/9512059}{{\ttfamily hep-th/9512059}}.

\bibitem{Billo:2016cpy}
M.~Billo, V.~Goncalves, E.~Lauria, and M.~Meineri, ``{Defects in conformal
  field theory},'' \href{http://dx.doi.org/10.1007/JHEP04(2016)091}{{\em JHEP}
  {\bfseries 04} (2016) 091}, \href{http://arxiv.org/abs/1601.02883}{{\ttfamily
  arXiv:1601.02883}}.

\bibitem{Bergshoeff:1987cm}
E.~Bergshoeff, E.~Sezgin, and P.~K. Townsend, ``{Supermembranes and
  eleven-dimensional supergravity},''
\href{http://dx.doi.org/10.1016/0370-2693(87)91272-X}{{\em Phys. Lett.}
  {\bfseries B189} (1987) 75--78}.

\bibitem{deWit:1998yu}
B.~de~Wit, K.~Peeters, J.~Plefka, and A.~Sevrin, ``{The M theory two-brane in
  AdS$_4\times S^7$ and AdS$_7\times S^4$},''
  \href{http://dx.doi.org/10.1016/S0370-2693(98)01340-9}{{\em Phys. Lett.}
  {\bfseries B443} (1998) 153--158},
\href{http://arxiv.org/abs/hep-th/9808052}{{\ttfamily hep-th/9808052}}.

\bibitem{Forste:1999yj}
S.~Forste, ``{Membrany corrections to the string anti-string potential in
  M5-brane theory},''
  \href{http://dx.doi.org/10.1088/1126-6708/1999/05/002}{{\em JHEP} {\bfseries
  05} (1999) 002},
\href{http://arxiv.org/abs/hep-th/9902068}{{\ttfamily hep-th/9902068}}.

\bibitem{Freund:1980xh}
P.~G.~O. Freund and M.~A. Rubin, ``{Dynamics of dimensional reduction},''
\href{http://dx.doi.org/10.1016/0370-2693(80)90590-0}{{\em Phys. Lett.}
  {\bfseries 97B} (1980) 233--235}.

\bibitem{Tseytlin:1999tp}
A.~A. Tseytlin and K.~Zarembo, ``{Magnetic interactions of D-branes and
  Wess-Zumino terms in superYang-Mills effective actions},''
  \href{http://dx.doi.org/10.1016/S0370-2693(99)01499-9}{{\em Phys. Lett.}
  {\bfseries B474} (2000) 95--102},
\href{http://arxiv.org/abs/hep-th/9911246}{{\ttfamily hep-th/9911246}}.

\bibitem{Gustavsson:2004dm}
A.~Gustavsson, ``{Dynamics of a wavy plane Wilson surface observable from
  AdS-CFT correspondence},''
  \href{http://dx.doi.org/10.1088/1126-6708/2005/01/022}{{\em JHEP} {\bfseries
  01} (2005) 022},
\href{http://arxiv.org/abs/hep-th/0411253}{{\ttfamily hep-th/0411253}}.

\bibitem{Liendo:2018ukf}
P.~Liendo, C.~Meneghelli, and V.~Mitev, ``{Bootstrapping the half-BPS line
  defect},'' \href{http://dx.doi.org/10.1007/JHEP10(2018)077}{{\em JHEP}
  {\bfseries 10} (2018) 077},
\href{http://arxiv.org/abs/1806.01862}{{\ttfamily arXiv:1806.01862}}.

\bibitem{Berman:2001fs}
D.~S. Berman and P.~Sundell, ``{AdS$_3$ OM theory and the selfdual string or
  membranes ending on the five -brane},''
  \href{http://dx.doi.org/10.1016/S0370-2693(02)01249-2}{{\em Phys. Lett.}
  {\bfseries B529} (2002) 171--177},
\href{http://arxiv.org/abs/hep-th/0105288}{{\ttfamily hep-th/0105288}}.

\bibitem{Mori:2014tca}
H.~Mori and S.~Yamaguchi, ``{M5-branes and Wilson surfaces in
  AdS$_{7}$/CFT$_{6}$ correspondence},''
  \href{http://dx.doi.org/10.1103/PhysRevD.90.026005}{{\em Phys. Rev.}
  {\bfseries D90} no.~2, (2014) 026005},
\href{http://arxiv.org/abs/1404.0930}{{\ttfamily arXiv:1404.0930}}.

\bibitem{DHoker:2008rje}
E.~D'Hoker, J.~Estes, M.~Gutperle, and D.~Krym, ``{Exact Half-BPS Flux
  Solutions in M-theory II: Global solutions asymptotic to AdS(7) x S4},''
  \href{http://dx.doi.org/10.1088/1126-6708/2008/12/044}{{\em JHEP} {\bfseries
  12} (2008) 044},
\href{http://arxiv.org/abs/0810.4647}{{\ttfamily arXiv:0810.4647 [hep-th]}}.

\bibitem{Lunin:2007ab}
O.~Lunin, ``{1/2-BPS states in M theory and defects in the dual CFTs},''
  \href{http://dx.doi.org/10.1088/1126-6708/2007/10/014}{{\em JHEP} {\bfseries
  10} (2007) 014},
\href{http://arxiv.org/abs/0704.3442}{{\ttfamily arXiv:0704.3442}}.

\bibitem{Chen:2007ir}
B.~Chen, W.~He, J.-B. Wu, and L.~Zhang, ``{M5-branes and Wilson surfaces},''
  \href{http://dx.doi.org/10.1088/1126-6708/2007/08/067}{{\em JHEP} {\bfseries
  08} (2007) 067},
\href{http://arxiv.org/abs/0707.3978}{{\ttfamily arXiv:0707.3978}}.

\bibitem{Rodgers:2018mvq}
R.~Rodgers, ``{Holographic entanglement entropy from probe M-theory branes},''
  \href{http://dx.doi.org/10.1007/JHEP03(2019)092}{{\em JHEP} {\bfseries 03}
  (2019) 092},
\href{http://arxiv.org/abs/1811.12375}{{\ttfamily arXiv:1811.12375}}.

\bibitem{Estes:2018tnu}
J.~Estes, D.~Krym, A.~O'Bannon, B.~Robinson, and R.~Rodgers, ``{Wilson surface
  central charge from holographic entanglement entropy},''
  \href{http://dx.doi.org/10.1007/JHEP05(2019)032}{{\em JHEP} {\bfseries 05}
  (2019) 032},
\href{http://arxiv.org/abs/1812.00923}{{\ttfamily arXiv:1812.00923}}.

\bibitem{Jensen:2018rxu}
K.~Jensen, A.~O'Bannon, B.~Robinson, and R.~Rodgers, ``{From the Weyl anomaly
  to entropy of two-dimensional boundaries and defects},''
  \href{http://dx.doi.org/10.1103/PhysRevLett.122.241602}{{\em Phys. Rev.
  Lett.} {\bfseries 122} no.~24, (2019) 241602},
\href{http://arxiv.org/abs/1812.08745}{{\ttfamily arXiv:1812.08745}}.

\bibitem{Alday:2007he}
L.~F. Alday and J.~Maldacena, ``{Comments on gluon scattering amplitudes via
  AdS/CFT},'' \href{http://dx.doi.org/10.1088/1126-6708/2007/11/068}{{\em JHEP}
  {\bfseries 11} (2007) 068},
\href{http://arxiv.org/abs/0710.1060}{{\ttfamily arXiv:0710.1060}}.

\bibitem{Polchinski:2011im}
J.~Polchinski and J.~Sully, ``{Wilson Loop Renormalization Group Flows},''
  \href{http://dx.doi.org/10.1007/JHEP10(2011)059}{{\em JHEP} {\bfseries 10}
  (2011) 059},
\href{http://arxiv.org/abs/1104.5077}{{\ttfamily arXiv:1104.5077}}.

\bibitem{Beccaria:2017rbe}
M.~Beccaria, S.~Giombi, and A.~Tseytlin, ``{Non-supersymmetric Wilson loop in
  ${\cal N}=4$ SYM and defect 1d CFT},''
\href{http://arxiv.org/abs/1712.06874}{{\ttfamily arXiv:1712.06874}}.

\bibitem{Beccaria:2019dws}
M.~Beccaria, S.~Giombi, and A.~A. Tseytlin, ``{Correlators on
  non-supersymmetric Wilson line in $ \mathcal{N}=4 $ SYM and
  AdS$_{2}$/CFT$_{1}$},'' \href{http://dx.doi.org/10.1007/JHEP05(2019)122}{{\em
  JHEP} {\bfseries 05} (2019) 122},
\href{http://arxiv.org/abs/1903.04365}{{\ttfamily arXiv:1903.04365}}.

\bibitem{Jensen:2015swa}
K.~Jensen and A.~O'Bannon, ``{Constraint on defect and boundary renormalization
  group flows},'' \href{http://dx.doi.org/10.1103/PhysRevLett.116.091601}{{\em
  Phys. Rev. Lett.} {\bfseries 116} no.~9, (2016) 091601},
\href{http://arxiv.org/abs/1509.02160}{{\ttfamily arXiv:1509.02160}}.

\bibitem{Casini:2018nym}
H.~Casini, I.~Salazar~Landea, and G.~Torroba, ``{Irreversibility in quantum
  field theories with boundaries},''
  \href{http://dx.doi.org/10.1007/JHEP04(2019)166}{{\em JHEP} {\bfseries 04}
  (2019) 166},
\href{http://arxiv.org/abs/1812.08183}{{\ttfamily arXiv:1812.08183}}.

\bibitem{Kobayashi:2018lil}
N.~Kobayashi, T.~Nishioka, Y.~Sato, and K.~Watanabe, ``{Towards a $C$-theorem
  in defect CFT},'' \href{http://dx.doi.org/10.1007/JHEP01(2019)039}{{\em JHEP}
  {\bfseries 01} (2019) 039}, \href{http://arxiv.org/abs/1810.06995}{{\ttfamily
  arXiv:1810.06995}}.

\bibitem{Drukker:2009sf}
N.~Drukker and J.~Plefka, ``{Superprotected $n$-point correlation functions of
  local operators in ${\cal N}=4$ super Yang-Mills},''
  \href{http://dx.doi.org/10.1088/1126-6708/2009/04/052}{{\em JHEP} {\bfseries
  04} (2009) 052},
\href{http://arxiv.org/abs/0901.3653}{{\ttfamily arXiv:0901.3653}}.

\bibitem{Tseytlin:1996hi}
A.~A. Tseytlin, ``{'No force' condition and BPS combinations of p-branes in
  eleven-dimensions and ten-dimensions},''
  \href{http://dx.doi.org/10.1016/S0550-3213(96)00692-X}{{\em Nucl. Phys.}
  {\bfseries B487} (1997) 141--154},
\href{http://arxiv.org/abs/hep-th/9609212}{{\ttfamily hep-th/9609212}}.

\bibitem{Klebanov:1996un}
I.~R. Klebanov and A.~A. Tseytlin, ``{Entropy of near extremal black
  $p$-branes},'' \href{http://dx.doi.org/10.1016/0550-3213(96)00295-7}{{\em
  Nucl. Phys.} {\bfseries B475} (1996) 164--178},
\href{http://arxiv.org/abs/hep-th/9604089}{{\ttfamily hep-th/9604089}}.

\bibitem{Tseytlin:2000sf}
A.~A. Tseytlin, ``{$R^4$ terms in 11 dimensions and conformal anomaly of (2,0)
  theory},'' \href{http://dx.doi.org/10.1016/S0550-3213(00)00380-1}{{\em Nucl.
  Phys.} {\bfseries B584} (2000) 233--250},
\href{http://arxiv.org/abs/hep-th/0005072}{{\ttfamily hep-th/0005072}}.

\bibitem{Intriligator:2000eq}
K.~A. Intriligator, ``{Anomaly matching and a Hopf-Wess-Zumino term in 6d,
  ${\cal N}=(2,0)$ field theories},''
  \href{http://dx.doi.org/10.1016/S0550-3213(00)00148-6}{{\em Nucl. Phys.}
  {\bfseries B581} (2000) 257--273},
\href{http://arxiv.org/abs/hep-th/0001205}{{\ttfamily hep-th/0001205}}.

\bibitem{Abanov:1999qz}
A.~G. Abanov and P.~B. Wiegmann, ``{Theta terms in nonlinear sigma models},''
  \href{http://dx.doi.org/10.1016/S0550-3213(99)00820-2}{{\em Nucl. Phys.}
  {\bfseries B570} (2000) 685--698},
\href{http://arxiv.org/abs/hep-th/9911025}{{\ttfamily hep-th/9911025}}.

\bibitem{Forste:1999qn}
S.~Forste, D.~Ghoshal, and S.~Theisen, ``{Stringy corrections to the Wilson
  loop in ${\cal N}=4$ superYang-Mills theory},''
  \href{http://dx.doi.org/10.1088/1126-6708/1999/08/013}{{\em JHEP} {\bfseries
  08} (1999) 013},
\href{http://arxiv.org/abs/hep-th/9903042}{{\ttfamily hep-th/9903042}}.

\bibitem{Buchbinder:2014nia}
E.~I. Buchbinder and A.~A. Tseytlin, ``{$1/N$ correction in the D3-brane
  description of a circular Wilson loop at strong coupling},''
  \href{http://dx.doi.org/10.1103/PhysRevD.89.126008}{{\em Phys. Rev.}
  {\bfseries D89} no.~12, (2014) 126008},
\href{http://arxiv.org/abs/1404.4952}{{\ttfamily arXiv:1404.4952}}.

\bibitem{Camporesi:1994ga}
R.~Camporesi and A.~Higuchi, ``{Spectral functions and zeta functions in
  hyperbolic spaces},''
\href{http://dx.doi.org/10.1063/1.530850}{{\em J. Math. Phys.} {\bfseries 35}
  (1994) 4217--4246}.

\bibitem{Giombi:2013fka}
S.~Giombi and I.~R. Klebanov, ``{One loop tests of higher spin AdS/CFT},''
  \href{http://dx.doi.org/10.1007/JHEP12(2013)068}{{\em JHEP} {\bfseries 12}
  (2013) 068},
\href{http://arxiv.org/abs/1308.2337}{{\ttfamily arXiv:1308.2337}}.

\bibitem{Giombi:2016pvg}
S.~Giombi, I.~R. Klebanov, and Z.~M. Tan, ``{The ABC of higher-spin AdS/CFT},''
  \href{http://dx.doi.org/10.3390/universe4010018}{{\em Universe} {\bfseries 4}
  no.~1, (2018) 18},
\href{http://arxiv.org/abs/1608.07611}{{\ttfamily arXiv:1608.07611}}.

\bibitem{Diaz:2007an}
D.~E. Diaz and H.~Dorn, ``{Partition functions and double-trace deformations in
  AdS/CFT},'' \href{http://dx.doi.org/10.1088/1126-6708/2007/05/046}{{\em JHEP}
  {\bfseries 05} (2007) 046},
\href{http://arxiv.org/abs/hep-th/0702163}{{\ttfamily hep-th/0702163}}.

\bibitem{Mezei:2018url}
M.~Mezei, S.~S. Pufu, and Y.~Wang, ``{Chern-Simons theory from M5-branes and
  calibrated M2-branes},''
  \href{http://dx.doi.org/10.1007/JHEP08(2019)165}{{\em JHEP} {\bfseries 08}
  (2019) 165}, \href{http://arxiv.org/abs/1812.07572}{{\ttfamily
  arXiv:1812.07572}}.

\bibitem{Schwimmer:2008yh}
A.~Schwimmer and S.~Theisen, ``{Entanglement entropy, trace anomalies and
  holography},'' \href{http://dx.doi.org/10.1016/j.nuclphysb.2008.04.015}{{\em
  Nucl. Phys.} {\bfseries B801} (2008) 1--24},
\href{http://arxiv.org/abs/0802.1017}{{\ttfamily arXiv:0802.1017}}.

\bibitem{graham:1999pm}
C.~R. Graham and E.~Witten, ``{Conformal anomaly of submanifold observables in
  $AdS$/CFT correspondence},''
  \href{http://dx.doi.org/10.1016/S0550-3213(99)00055-3}{{\em Nucl. Phys.}
  {\bfseries B546} (1999) 52--64},
  \href{http://arxiv.org/abs/hep-th/9901021}{{\ttfamily hep-th/9901021}}.

\bibitem{Bianchi:2019sxz}
L.~Bianchi and M.~Lemos, ``{Superconformal surfaces in four dimensions},''
  \href{http://dx.doi.org/10.1007/JHEP06(2020)056}{{\em JHEP} {\bfseries 06}
  (2020) 056}, \href{http://arxiv.org/abs/1911.05082}{{\ttfamily
  arXiv:1911.05082 [hep-th]}}.

\bibitem{Lewkowycz:2013laa}
A.~Lewkowycz and J.~Maldacena, ``{Exact results for the entanglement entropy
  and the energy radiated by a quark},''
  \href{http://dx.doi.org/10.1007/JHEP05(2014)025}{{\em JHEP} {\bfseries 05}
  (2014) 025},
\href{http://arxiv.org/abs/1312.5682}{{\ttfamily arXiv:1312.5682}}.

\bibitem{bianchi:2018zpb}
L.~Bianchi, M.~Lemos, and M.~Meineri, ``{Line defects and radiation in
  $\mathcal{N}=2$ conformal theories},''
  \href{http://dx.doi.org/10.1103/PhysRevLett.121.141601}{{\em Phys. Rev.
  Lett.} {\bfseries 121} no.~14, (2018) 141601},
\href{http://arxiv.org/abs/1805.04111}{{\ttfamily arXiv:1805.04111}}.

\bibitem{Drukker:2020atp}
N.~Drukker, M.~Probst, and M.~Tr\'epanier, ``{Defect CFT techniques in the 6d
  $\mathcal{N} = (2,0)$ theory},''
  \href{http://dx.doi.org/10.1007/JHEP03(2021)261}{{\em JHEP} {\bfseries 03}
  (2021) 261}, \href{http://arxiv.org/abs/2009.10732}{{\ttfamily
  arXiv:2009.10732 [hep-th]}}.

\bibitem{Chalabi:2020iie}
A.~Chalabi, A.~O'Bannon, B.~Robinson, and J.~Sisti, ``{Central charges of 2d
  superconformal defects},''
  \href{http://dx.doi.org/10.1007/JHEP05(2020)095}{{\em JHEP} {\bfseries 05}
  (2020) 095}, \href{http://arxiv.org/abs/2003.02857}{{\ttfamily
  arXiv:2003.02857 [hep-th]}}.

\bibitem{Beccaria:2015ypa}
M.~Beccaria and A.~A. Tseytlin, ``{Conformal anomaly \ c-coefficients of
  superconformal 6d theories},''
  \href{http://dx.doi.org/10.1007/JHEP01(2016)001}{{\em JHEP} {\bfseries 01}
  (2016) 001},
\href{http://arxiv.org/abs/1510.02685}{{\ttfamily arXiv:1510.02685}}.

\bibitem{Henningson:1998gx}
M.~Henningson and K.~Skenderis, ``{The Holographic Weyl anomaly},''
  \href{http://dx.doi.org/10.1088/1126-6708/1998/07/023}{{\em JHEP} {\bfseries
  07} (1998) 023},
\href{http://arxiv.org/abs/hep-th/9806087}{{\ttfamily hep-th/9806087}}.

\bibitem{Beccaria:2014qea}
M.~Beccaria, G.~Macorini, and A.~A. Tseytlin, ``{Supergravity one-loop
  corrections on AdS$_7$ and AdS$_3$, higher spins and AdS/CFT},''
  \href{http://dx.doi.org/10.1016/j.nuclphysb.2015.01.014}{{\em Nucl. Phys.}
  {\bfseries B892} (2015) 211--238},
\href{http://arxiv.org/abs/1412.0489}{{\ttfamily arXiv:1412.0489}}.

\bibitem{Mansfield:2003bg}
P.~Mansfield, D.~Nolland, and T.~Ueno, ``{Order $1 / N^3$ corrections to the
  conformal anomaly of the (2,0) theory in six-dimensions},''
  \href{http://dx.doi.org/10.1016/S0370-2693(03)00777-9}{{\em Phys. Lett.}
  {\bfseries B566} (2003) 157--163},
\href{http://arxiv.org/abs/hep-th/0305015}{{\ttfamily hep-th/0305015}}.

\bibitem{Beem:2014kka}
C.~Beem, L.~Rastelli, and B.~C. van Rees, ``{$ \mathcal{W} $ symmetry in six
  dimensions},'' \href{http://dx.doi.org/10.1007/JHEP05(2015)017}{{\em JHEP}
  {\bfseries 05} (2015) 017},
\href{http://arxiv.org/abs/1404.1079}{{\ttfamily arXiv:1404.1079}}.

\bibitem{Ohmori:2014kda}
K.~Ohmori, H.~Shimizu, Y.~Tachikawa, and K.~Yonekura, ``{Anomaly polynomial of
  general 6d SCFTs},'' \href{http://dx.doi.org/10.1093/ptep/ptu140}{{\em PTEP}
  {\bfseries 2014} no.~10, (2014) 103B07},
\href{http://arxiv.org/abs/1408.5572}{{\ttfamily arXiv:1408.5572}}.

\bibitem{Giombi:2013yva}
S.~Giombi, I.~R. Klebanov, S.~S. Pufu, B.~R. Safdi, and G.~Tarnopolsky, ``{AdS
  description of induced higher-spin gauge theory},''
  \href{http://dx.doi.org/10.1007/JHEP10(2013)016}{{\em JHEP} {\bfseries 10}
  (2013) 016},
\href{http://arxiv.org/abs/1306.5242}{{\ttfamily arXiv:1306.5242}}.

\bibitem{Beccaria:2014jxa}
M.~Beccaria, X.~Bekaert, and A.~A. Tseytlin, ``{Partition function of free
  conformal higher spin theory},''
  \href{http://dx.doi.org/10.1007/JHEP08(2014)113}{{\em JHEP} {\bfseries 08}
  (2014) 113},
\href{http://arxiv.org/abs/1406.3542}{{\ttfamily arXiv:1406.3542}}.

\bibitem{Vassilevich:2003xt}
D.~V. Vassilevich, ``{Heat kernel expansion: User's manual},''
  \href{http://dx.doi.org/10.1016/j.physrep.2003.09.002}{{\em Phys. Rept.}
  {\bfseries 388} (2003) 279--360},
\href{http://arxiv.org/abs/hep-th/0306138}{{\ttfamily hep-th/0306138}}.

\bibitem{Freedman:1998tz}
D.~Z. Freedman, S.~D. Mathur, A.~Matusis, and L.~Rastelli, ``{Correlation
  functions in the CFT$_d$/AdS$_{d+1}$ correspondence},''
  \href{http://dx.doi.org/10.1016/S0550-3213(99)00053-X}{{\em Nucl. Phys.}
  {\bfseries B546} (1999) 96--118},
\href{http://arxiv.org/abs/hep-th/9804058}{{\ttfamily hep-th/9804058}}.

\bibitem{Drukker:2011za}
N.~Drukker and V.~Forini, ``{Generalized quark-antiquark potential at weak and
  strong coupling},'' \href{http://dx.doi.org/10.1007/JHEP06(2011)131}{{\em
  JHEP} {\bfseries 06} (2011) 131},
\href{http://arxiv.org/abs/1105.5144}{{\ttfamily arXiv:1105.5144}}.

\bibitem{Correa:2012at}
D.~Correa, J.~Henn, J.~Maldacena, and A.~Sever, ``{An exact formula for the
  radiation of a moving quark in ${\cal N}=4$ super Yang Mills},''
  \href{http://dx.doi.org/10.1007/JHEP06(2012)048}{{\em JHEP} {\bfseries 06}
  (2012) 048},
\href{http://arxiv.org/abs/1202.4455}{{\ttfamily arXiv:1202.4455}}.

\bibitem{bianchi:2015liz}
L.~Bianchi, M.~Meineri, R.~C. Myers, and M.~Smolkin, ``{R\'enyi entropy and
  conformal defects},'' \href{http://dx.doi.org/10.1007/JHEP07(2016)076}{{\em
  JHEP} {\bfseries 07} (2016) 76},
  \href{http://arxiv.org/abs/1511.06713}{{\ttfamily arXiv:1511.06713}}.

\bibitem{Rastelli:2019gtj}
L.~Rastelli, K.~Roumpedakis, and X.~Zhou, ``{$AdS_3\times S^3$ tree-level
  correlators: Hidden six-dimensional conformal symmetry},''
  \href{http://dx.doi.org/10.1007/JHEP10(2019)140}{{\em JHEP} {\bfseries 10}
  (2019) 140},
\href{http://arxiv.org/abs/1905.11983}{{\ttfamily arXiv:1905.11983}}.

\bibitem{Dolan:2004mu}
F.~A. Dolan, L.~Gallot, and E.~Sokatchev, ``{On four-point functions of 1/2-BPS
  operators in general dimensions},''
  \href{http://dx.doi.org/10.1088/1126-6708/2004/09/056}{{\em JHEP} {\bfseries
  09} (2004) 056},
\href{http://arxiv.org/abs/hep-th/0405180}{{\ttfamily hep-th/0405180}}.

\bibitem{Arutyunov:2002fh}
G.~Arutyunov, F.~A. Dolan, H.~Osborn, and E.~Sokatchev, ``{Correlation
  functions and massive Kaluza-Klein modes in the AdS/CFT correspondence},''
  \href{http://dx.doi.org/10.1016/S0550-3213(03)00448-6}{{\em Nucl. Phys.}
  {\bfseries B665} (2003) 273--324},
\href{http://arxiv.org/abs/hep-th/0212116}{{\ttfamily hep-th/0212116}}.

\bibitem{Mack:2009mi}
G.~Mack, ``{D-independent representation of conformal rield theories in $D$
  dimensions via transformation to auxiliary dual resonance models. Scalar
  amplitudes},''
\href{http://arxiv.org/abs/0907.2407}{{\ttfamily arXiv:0907.2407}}.

\bibitem{Penedones:2010ue}
J.~Penedones, ``{Writing CFT correlation functions as AdS scattering
  amplitudes},'' \href{http://dx.doi.org/10.1007/JHEP03(2011)025}{{\em JHEP}
  {\bfseries 03} (2011) 025},
\href{http://arxiv.org/abs/1011.1485}{{\ttfamily arXiv:1011.1485}}.

\bibitem{Zhou:2018sfz}
X.~Zhou, ``{Recursion relations in Witten diagrams and conformal partial
  waves},'' \href{http://dx.doi.org/10.1007/JHEP05(2019)006}{{\em JHEP}
  {\bfseries 05} (2019) 006},
\href{http://arxiv.org/abs/1812.01006}{{\ttfamily arXiv:1812.01006}}.

\bibitem{Heemskerk:2009pn}
I.~Heemskerk, J.~Penedones, J.~Polchinski, and J.~Sully, ``{Holography from
  conformal field theory},''
  \href{http://dx.doi.org/10.1088/1126-6708/2009/10/079}{{\em JHEP} {\bfseries
  10} (2009) 079},
\href{http://arxiv.org/abs/0907.0151}{{\ttfamily arXiv:0907.0151}}.

\bibitem{Kac:1977qb}
V.~G. Kac, ``{A Sketch of Lie superalgebra theory},''
\href{http://dx.doi.org/10.1007/BF01609166}{{\em Commun. Math. Phys.}
  {\bfseries 53} (1977) 31--64}.

\bibitem{Frappat:1996pb}
L.~Frappat, P.~Sorba, and A.~Sciarrino, ``{Dictionary on Lie superalgebras},''
\href{http://arxiv.org/abs/hep-th/9607161}{{\ttfamily hep-th/9607161}}.

\bibitem{Gimenez-Grau:2019hez}
A.~Gimenez-Grau and P.~Liendo, ``{Bootstrapping line defects in $\mathcal{N}=2$
  theories},'' \href{http://dx.doi.org/10.1007/JHEP03(2020)121}{{\em JHEP}
  {\bfseries 03} (2020) 121}, \href{http://arxiv.org/abs/1907.04345}{{\ttfamily
  arXiv:1907.04345 [hep-th]}}.

\bibitem{Alday:2019nin}
L.~F. Alday and X.~Zhou, ``{Simplicity of AdS Supergravity at One Loop},''
  \href{http://dx.doi.org/10.1007/JHEP09(2020)008}{{\em JHEP} {\bfseries 09}
  (2020) 008}, \href{http://arxiv.org/abs/1912.02663}{{\ttfamily
  arXiv:1912.02663 [hep-th]}}.

\bibitem{Eden:2000bk}
B.~Eden, A.~C. Petkou, C.~Schubert, and E.~Sokatchev, ``{Partial
  nonrenormalization of the stress tensor four point function in ${\cal N}=4$
  SYM and AdS/CFT},''
  \href{http://dx.doi.org/10.1016/S0550-3213(01)00151-1}{{\em Nucl. Phys.}
  {\bfseries B607} (2001) 191--212},
\href{http://arxiv.org/abs/hep-th/0009106}{{\ttfamily hep-th/0009106}}.

\bibitem{Nirschl:2004pa}
M.~Nirschl and H.~Osborn, ``{Superconformal Ward identities and their
  solution},'' \href{http://dx.doi.org/10.1016/j.nuclphysb.2005.01.013}{{\em
  Nucl. Phys.} {\bfseries B711} (2005) 409--479},
\href{http://arxiv.org/abs/hep-th/0407060}{{\ttfamily hep-th/0407060}}.

\bibitem{Arutyunov:2000py}
G.~Arutyunov and S.~Frolov, ``{Four point functions of lowest weight CPOs in
  ${\cal N}=4$ SYM$_4$ in supergravity approximation},''
  \href{http://dx.doi.org/10.1103/PhysRevD.62.064016}{{\em Phys. Rev.}
  {\bfseries D62} (2000) 064016},
\href{http://arxiv.org/abs/hep-th/0002170}{{\ttfamily hep-th/0002170}}.

\bibitem{Caron-Huot:2018kta}
S.~Caron-Huot and A.-K. Trinh, ``{All tree-level correlators in AdS$_{5}\times
  S_{5}$ supergravity: hidden ten-dimensional conformal symmetry},''
  \href{http://dx.doi.org/10.1007/JHEP01(2019)196}{{\em JHEP} {\bfseries 01}
  (2019) 196},
\href{http://arxiv.org/abs/1809.09173}{{\ttfamily arXiv:1809.09173}}.

\bibitem{Rastelli:2016nze}
L.~Rastelli and X.~Zhou, ``{Mellin amplitudes for $AdS_5\times S^5$},''
  \href{http://dx.doi.org/10.1103/PhysRevLett.118.091602}{{\em Phys. Rev.
  Lett.} {\bfseries 118} no.~9, (2017) 091602},
\href{http://arxiv.org/abs/1608.06624}{{\ttfamily arXiv:1608.06624}}.

\bibitem{Rastelli:2017udc}
L.~Rastelli and X.~Zhou, ``{How to succeed at holographic correlators without
  really trying},'' \href{http://dx.doi.org/10.1007/JHEP04(2018)014}{{\em JHEP}
  {\bfseries 04} (2018) 014},
\href{http://arxiv.org/abs/1710.05923}{{\ttfamily arXiv:1710.05923}}.

\bibitem{Rastelli:2017ymc}
L.~Rastelli and X.~Zhou, ``{Holographic four-point functions in the (2, 0)
  theory},'' \href{http://dx.doi.org/10.1007/JHEP06(2018)087}{{\em JHEP}
  {\bfseries 06} (2018) 087},
\href{http://arxiv.org/abs/1712.02788}{{\ttfamily arXiv:1712.02788}}.

\bibitem{Zhou:2017zaw}
X.~Zhou, ``{On superconformal four-point Mellin amplitudes in dimension
  $d>2$},'' \href{http://dx.doi.org/10.1007/JHEP08(2018)187}{{\em JHEP}
  {\bfseries 08} (2018) 187},
\href{http://arxiv.org/abs/1712.02800}{{\ttfamily arXiv:1712.02800}}.

\bibitem{Giombi:2018qox}
S.~Giombi and S.~Komatsu, ``{Exact Correlators on the Wilson Loop in
  $\mathcal{N}=4$ SYM: Localization, Defect CFT, and Integrability},''
  \href{http://dx.doi.org/10.1007/JHEP05(2018)109}{{\em JHEP} {\bfseries 05}
  (2018) 109}, \href{http://arxiv.org/abs/1802.05201}{{\ttfamily
  arXiv:1802.05201 [hep-th]}}. [Erratum: JHEP 11, 123 (2018)].

\bibitem{Giombi:2018hsx}
S.~Giombi and S.~Komatsu, ``{More Exact Results in the Wilson Loop Defect CFT:
  Bulk-Defect OPE, Nonplanar Corrections and Quantum Spectral Curve},''
  \href{http://dx.doi.org/10.1088/1751-8121/ab046c}{{\em J.\ Phys.\ A}
  {\bfseries 52} no.~12, (2019) 125401},
  \href{http://arxiv.org/abs/1811.02369}{{\ttfamily arXiv:1811.02369
  [hep-th]}}.

\end{thebibliography}\endgroup
\end{document}